\documentclass[a4paper,11pt]{article}
\usepackage{pos}
\usepackage[utf8]{inputenc}

\hypersetup{%
	pdftitle    = {Muon g-2: Lattice calculations of the hadronic vacuum polarization},
	pdfauthor   = {S.~Kuberski},
}%

\title{Muon $g-2$: Lattice calculations of the hadronic vacuum polarization}

\author[]{Simon Kuberski}

\affiliation{Theoretical Physics Department, CERN, 1211 Geneva 23, Switzerland}

\affiliation{Helmholtz-Institut Mainz, Johannes Gutenberg-Universit\"at
	Mainz, Germany}

\affiliation{GSI Helmholtz Centre for Heavy Ion Research, Darmstadt,
Germany}

\emailAdd{simon.kuberski@cern.ch}

\abstract{
The experimental uncertainty on the anomalous magnetic moment of the muon has been significantly reduced with the recent results of the Fermilab $g-2$ experiment, and a further reduction is expected in the near future. The precision of the Standard Model prediction needs to improve correspondingly to increase the sensitivity of tests for physics beyond the Standard Model. The largest uncertainty is due to contributions from the strong interaction, in particular the hadronic vacuum polarization (HVP) contribution.

Lattice QCD calculations have the potential to provide precise ab initio predictions of the HVP contribution. We review the state of lattice QCD calculations, focusing on the dominant sources of uncertainty that need to be controlled to provide results with sub-percent precision.

\vspace{3cm}\hfill\textit{ CERN-TH-2023-244, MITP-23-079}
}

\FullConference{The 40th International Symposium on Lattice Field Theory (Lattice 2023)\\
July 31st - August 4th, 2023\\
Fermi National Accelerator Laboratory\\}

\newcommand{\amu}{a_\mu^{\mathrm{hvp}}}
\newcommand{\ahvp}{a_\mu^{\rm hvp}}
\newcommand{\awin}{(a_\mu^{\mathrm{hvp}})^{\rm ID}}
\newcommand{\awinl}{(a_\mu^{\mathrm{hvp}})^{\rm ID, l}}
\newcommand{\asd}{(a_\mu^{\mathrm{hvp}})^{\rm SD}}
\newcommand{\asdl}{(a_\mu^{\mathrm{hvp}})^{\rm SD, l}}
\newcommand{\ald}{(a_\mu^{\mathrm{hvp}})^{\rm LD}}
\newcommand{\ahlbl}{a_\mu^{\rm hlbl}}
\begin{document}
\maketitle

\section{Introduction}
A long-standing tension between the Standard Model (SM) prediction and the
experimentally measured value of the anomalous magnetic moment of the 
muon $a_\mu$ has sparked both theoretical and experimental efforts, hoping to 
enhance the significance of the tension in the search for  
physics beyond the Standard Model of particle physics.

The anomalous magnetic moment of a lepton $l$ is defined as the 
deviation of its $g$-factor from its classical value $g=2$,
\begin{align}
	a_l = \frac{1}{2}(g_l- 2)\,,
\end{align}
due to quantum loop corrections from QED, electroweak and strong interactions
and possibly due to physics beyond the Standard Model. 
Contributions to $a_l$ induced by heavy particles with mass $M$ 
which are not part of the Standard Model enter proportionally to
$m_l^2/M^2$ and are thus enhanced by four orders of magnitude, when 
considering $a_\mu$ instead of $a_e$.
The anomalous
magnetic moments of the electron and the muon have been determined
to very high precision, i.e., to 0.11\,ppb \cite{Fan:2022eto} 
and 0.19\,ppm \cite{Bennett:2006fi, Muong-2:2021ojo, Muong-2:2023cdq},
respectively. 
The uncertainty of $a_\mu$ is expected to reduce further when the data 
from Runs 4-6 of the Fermilab $g-2$ experiment will be included in the
experimental average.

To test for physics beyond the Standard Model, SM predictions have to be
known at the same level of precision. Whereas QED 
\cite{Aoyama:2012wk, Aoyama:2019ryr}
and electroweak 
\cite{Czarnecki:2002nt, Gnendiger:2013pva} 
contributions to $a_\mu$ 
are known to a precision that exceeds the experimental one by far,
the uncertainty of the leading-order hadronic contributions to $a_\mu$ dominate
the theory uncertainty. These are the hadronic vacuum polarization (HVP) 
contribution $\ahvp$,
entering at order $\alpha^2$, and the hadronic light-by-light scattering
(HLBL) contribution $\ahlbl$ at $\mathrm{O}(\alpha^3)$. 
Whereas, thanks to recent work, SM predictions of $\ahlbl$ approach the
precision target of $10\%$, also in the lattice calculations of
\cite{Blum:2019ugy, Chao:2021tvp, Chao:2022xzg, Blum:2023vlm},
the situation is less favorable in the case of $\ahvp$, where a target
uncertainty of about 0.2\% will be needed to make full use of the advancements
on the experimental side.

Traditionally, $\ahvp$ has been computed from experimental data for the 
cross-section $\sigma(e^+e^-\to \text{hadrons})$ and a dispersion
relation via,
\begin{align}
	\ahvp = \left(\frac{\alpha m_\mu}{3\pi}\right)^2 \int_{m_{\pi^0}^2}^\infty \mathrm{d}s\frac{\hat{K}(s)}{s^2} R(s)\,,\qquad R(s) = \frac{\sigma(e^+e^-\to \text{hadrons})}{4\pi \alpha ^2 / (3s)}\,,
\end{align}
where $\hat{K}(s)$ is a known QED kernel function \cite{Brodsky:1967sr}. 
The precision of the evaluation in the White Paper of the Muon $g-2$
theory initiative \cite{Aoyama:2020ynm} based on
\cite{Davier:2017zfy,Keshavarzi:2018mgv,Colangelo:2018mtw,Hoferichter:2019mqg,Davier:2019can,Keshavarzi:2019abf,Kurz:2014wya}
is limited by a tension between the two most precise experimental data 
sets by the KLOE and BABAR experiments in the $e^+e^-\to\pi^+\pi^-$ channel
that contributes dominantly to $\ahvp$. 
A recent new measurement of this contribution with the CMD-3 detector 
\cite{CMD-3:2023alj}
is in tension with all previously obtained results and would potentially
lead to an inflation of uncertainties, when included in the analysis.
An alternative determination based on $\mu e$ scattering at the MUonE 
experiment  \cite{Abbiendi:2016xup} is aiming at a precision at the level
of $0.3\%$. The data taking can only be started after the long shutdown 3
at CERN in 2029.

Lattice QCD provides the optimal framework to compute the hadronic 
contributions to $a_\mu$ from first principles. Recent work has shown
that precise results can be obtained, keeping all sources of uncertainty
under control. In the following, we briefly recapitulate how $\ahvp$ is 
computed from lattice QCD and collect recent results for $\ahvp$ and several
sub-contributions. We point out the dominant and subleading sources of
uncertainty and how they are addressed in modern calculations.

\section{The HVP contribution from lattice QCD \label{s:hvp_lat}}

\begin{figure}
	\centering
	$\vcenter{\hbox{\includegraphics[width=.49\textwidth]{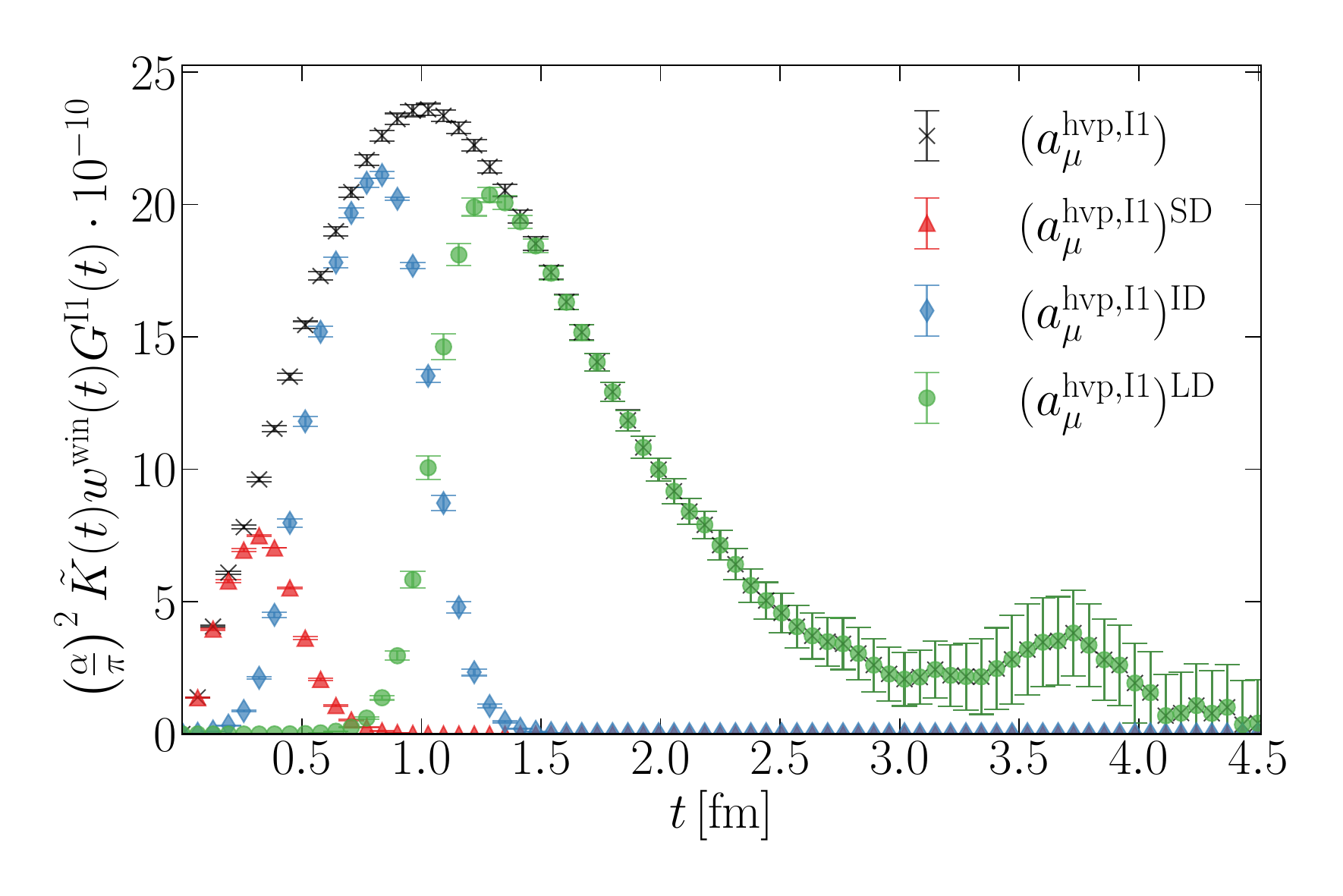}}}$
	$\vcenter{\hbox{\includegraphics[width=.49\textwidth]{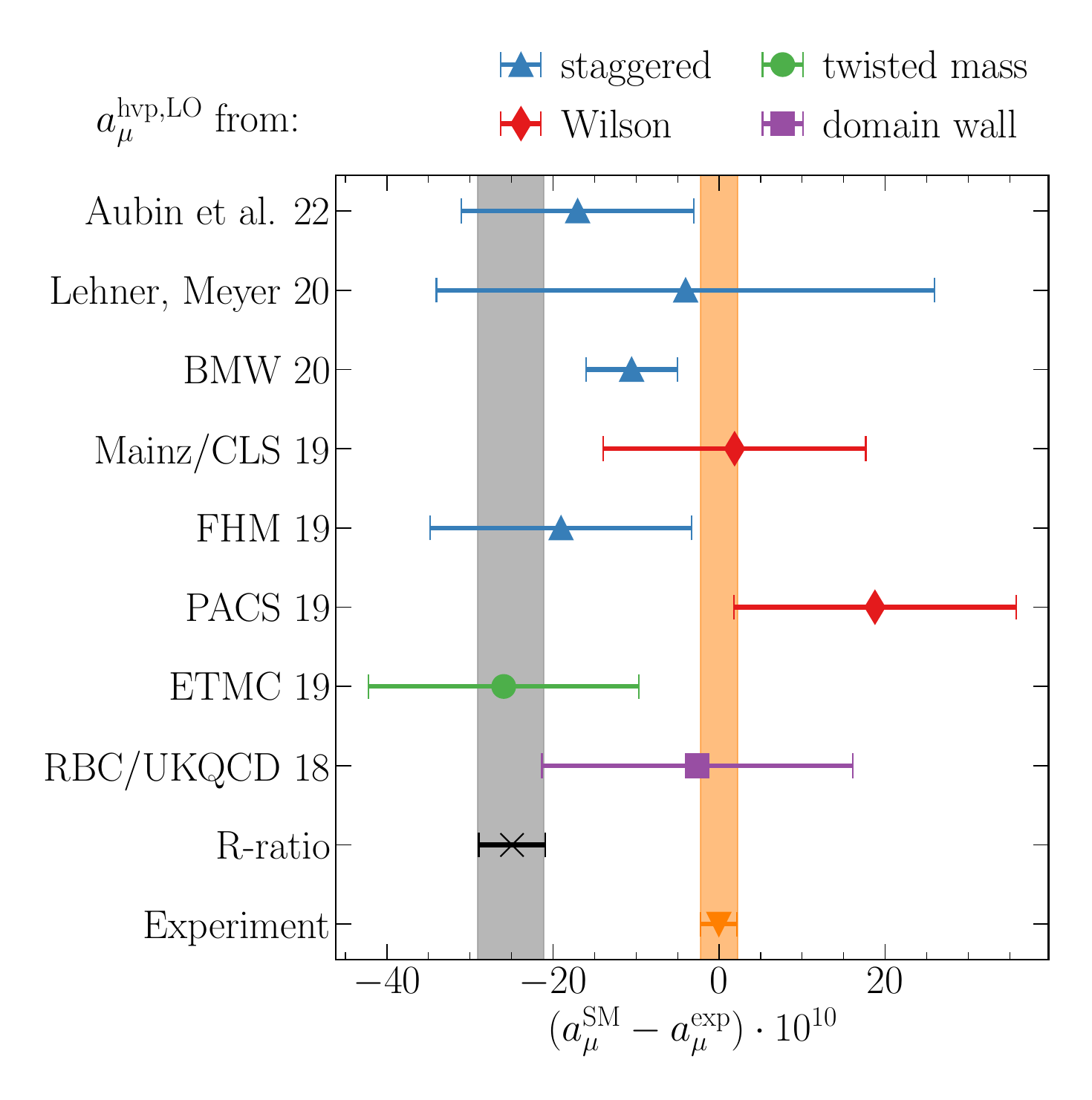}}}$
	\caption{\label{f:TMR_integrand}\label{f:comp_amu}
		Left: The integrand of eq.~(\ref{e:def_TMR}) for the isovector correlation function on an ensemble at physical pion mass \cite{Mohler:2017wnb}. The black crosses show the full integrand whereas the colored data show the integrands of the three windows observables, based on eq.~(\ref{e:def_windows}).
		Right: Comparison of SM predictions for $a_\mu$ with the experimental world average \cite{Bennett:2006fi, Muong-2:2021ojo, Muong-2:2023cdq}. The SM predictions differ only in the leading-order HVP contribution from the R-ratio \cite{Davier:2017zfy,Keshavarzi:2018mgv,Colangelo:2018mtw,Hoferichter:2019mqg,Davier:2019can,Keshavarzi:2019abf,Kurz:2014wya} or lattice QCD 
		\cite{Blum:2018mom, Giusti:2019xct, Shintani:2019wai, Davies:2019efs, Gerardin:2019rua, Borsanyi:2020mff, Lehner:2020crt, Aubin:2022hgm}. The lattice QCD results are grouped by fermion discretization.
	}
\end{figure}
The natural starting point for the computation of the HVP in Euclidean 
space-time is the polarization tensor,
\begin{align} \label{e:def_pi}
	\Pi_{\mu\nu}(Q) = \int \mathrm{d}^4x\, \mathrm{e}^{iQ\cdot x}\langle j_\mu^{\rm em}(x)\,j_\nu^{\rm em}(0) \rangle = (Q_\mu Q_\nu - \delta_{\mu\nu} Q^2) \Pi(Q^2)\,,
\end{align}
based on the two-point function of the hadronic part of the electromagnetic
current,
\begin{align}\label{e:def_jmu}
	j_\mu^{\rm em}=\frac{2}{3}\bar{u}\gamma_\mu u
	-\frac{1}{3}\bar{d}\gamma_\mu d
	-\frac{1}{3}\bar{s}\gamma_\mu s
	+\frac{2}{3}\bar{c}\gamma_\mu c
	-\frac{1}{3}\bar{b}\gamma_\mu b
	+\frac{2}{3}\bar{t}\gamma_\mu t\,.
\end{align}
As pointed out in \cite{Blum:2002ii}, a lattice calculation of the momentum
dependent hadronic tensor allows one to compute the leading-order HVP 
contribution to the muon $g-2$
via
\begin{align}\label{e:amu_blum}
	\amu = \left(\frac{\alpha}{\pi}\right)^2 \int_{0}^{\infty}\mathrm{d}Q^2f(Q^2)\hat{\Pi}(Q^2)\,,\qquad \text{with}\quad \hat{\Pi}(Q^2) = 4\pi^2 \left[\Pi(Q^2) - \Pi(0)\right]
\end{align}
where $f(Q^2)$ is a known analytic function that encodes (infinite volume)
QED and depends on the lepton mass $m_\mu$.
The most recent lattice calculations make use of the time-momentum
representation (TMR) to compute $\amu$. As first shown in 
\cite{Bernecker:2011gh} the subtracted vacuum polarization function $\hat{\Pi}$
may be written in terms of the spatially-summed, zero-momentum vector 
two-point correlation function 
\begin{align}\label{e:def_G}
	G(t)=-\frac{a^3}{3}\sum_{k=1}^3\sum_{\vec{x}}\left\langle
	j_k^{\rm em}(t,\vec{x})\,j_k^{\rm em}(0) \right\rangle\,,
\end{align}
via
\begin{align}\label{e:def_Pisub}
	\hat{\Pi}(Q^2) = \frac{4\pi^2}{Q^2}\int_0^\infty \mathrm{d}t\,G(t) \left[Q^2t^2 - 4\sin^2\left(\frac{1}{2} Q t\right)\right]\,.
\end{align}
By switching the order of integration over imaginary time and momentum in
eq.~(\ref{e:amu_blum}), one obtains
\begin{align}\label{e:def_TMR}
	\ahvp := \left(\frac{\alpha}{\pi}\right)^2\int_0^{\infty} dt
	\,G(t)\widetilde{K}(t)\,,
\end{align}
with an analytic QED kernel function $\widetilde{K}(t)$ 
\cite{DellaMorte:2017dyu} that gives weight to the long-distance 
regime of the correlation function.

Performing the Wick contractions of the two currents in eq.~(\ref{e:def_G})
gives rise to quark-connected and quark-disconnected correlation functions.
In isospin-symmetric QCD, these are conventionally written as
\begin{align}\label{e:Gt_flavdecomp}
	G(t) 
	= \frac{5}{9} G_{\rm l}(t) 
	+ \frac{1}{9} G_{\rm s}(t)
	+ \frac{4}{9} G_{\rm c}(t)
	+ \frac{1}{9} G_{\rm b}(t)
	+ G_{\rm disc}(t)\,.
\end{align}
The top quark contribution is neglected here because it is not accessible on
the lattice. It can be computed in perturbation theory but is too small to be
relevant at the current level of uncertainty. There exists one lattice result
\cite{Colquhoun:2014ica} for the bottom quark contribution, to which
perturbation theory is also applicable.
The term $G_{\rm disc}(t)$ collects all quark-disconnected contributions. 
In the decomposition into isovector and isoscalar contributions,
\begin{align}\label{e:Gt_isodecomp}
	G^{\rm I1}(t) = \frac{1}{2} G_{\rm l}(t)\,,\qquad 
	G^{\rm I0}(t)
	= \frac{1}{18} G_{\rm l}(t) 
	+ \frac{1}{9} G_{\rm s}(t)
	+ \frac{4}{9} G_{\rm c}(t)
	+ \frac{1}{9} G_{\rm b}(t)
	+ G_{\rm disc}(t)\,,
\end{align}
the finite-volume effects of light-connected and disconnected 
contributions largely cancel within $G^{\rm I0}$ \cite{Gerardin:2019rua},
facilitating the overall correction for these effects.

The typical shape of the integrand in eq.~(\ref{e:def_TMR}) for the isovector contribution
at physical value of the light quark mass is displayed by the black crosses
in Figure~\ref{f:TMR_integrand}. Whereas the correlation function decreases
exponentially with increasing time separation $t$, the QED kernel function
$\tilde{K}$ increases polynomially \cite{DellaMorte:2017dyu}. 
The interplay of the two results in a large contribution of time separations
beyond $1.5\,$fm, complicating the precise determination of the integral
due to a signal-to-noise problem in $G^{\rm I1}(t)$. The contribution from 
times beyond 4\,fm is numerically small. 
At the same time, the integrand probes very short distances where the
correlator may be determined precisely but is affected by significant cutoff
effects. 

In \cite{Bernecker:2011gh}, it has been suggested to split the integration 
in three different contributions at short, intermediate and long distances.
Smooth window functions in Euclidean time, defined by
\begin{align}\label{e:def_windows}
	{w^\mathrm{SD}}(t; t_0; t_1) &= [1 - \Theta(t, t_0, \Delta)]\,,
	&
	{w^\mathrm{ID}}(t; t_0; t_1) &=[\Theta(t,t_0,\Delta)-\Theta(t,t_1,\Delta)]\,,
	\nonumber
	\\
	{w^\mathrm{LD}}(t; t_0; t_1) &=  \Theta(t, t_1, \Delta)\,,
	&
	\Theta(t,t^\prime,\Delta)&:={\textstyle\frac{1}{2}}\left(
	1+\tanh[(t-t^\prime)/\Delta]\right)
\end{align}
with the parameters $t_0=0.4\,{\rm fm}$, $t_1=1.0\,{\rm fm}$ and
$\Delta=0.15\,{\rm fm}$ have been introduced in \cite{Blum:2018mom} 
to smoothly separate the three regions. So-called window observables
can be defined by inserting one of these functions $w^X(t; t_0; t_1)$ 
in the integrand of eq.~(\ref{e:def_TMR}). 
The resulting integrands add up to the integrand
of eq.~(\ref{e:def_TMR}) and are shown by colored markers
in Figure~\ref{f:TMR_integrand}. 
The window observables aid in untangling and confirming control over various
sources of systematic uncertainty, both at short and long distances.
Recent result the for short and intermediate
distance contributions will be discussed in section~\ref{s:windows}.

\section{Results for \texorpdfstring{$\ahvp$}{the HVP contribution} \label{s:results}}
The right panel of Figure~\ref{f:comp_amu} compares Standard Model predictions for the
leading-order HVP contribution to $a_\mu$ with the current world 
average for the experimentally determined $a_\mu^{\rm exp}$ from
\cite{Bennett:2006fi, Muong-2:2021ojo, Muong-2:2023cdq}. 
The black cross and gray error band denote the White Paper average 
from \cite{Aoyama:2020ynm}, where $\ahvp$ has been evaluated
using data-driven dispersive methods from the R-ratio and perturbative QCD
\cite{Davier:2017zfy,Keshavarzi:2018mgv,Colangelo:2018mtw,Hoferichter:2019mqg,Davier:2019can,Keshavarzi:2019abf,Kurz:2014wya}.
A $5.1\,\sigma$ tension is found when comparing this result with
$a_\mu^{\rm exp}$.

The colored data points are obtained by using a lattice QCD result for 
$a_\mu^{\rm hvp}$, instead of the White Paper average. They show
the most recent results of eight collaborations
\cite{Blum:2018mom, Giusti:2019xct, Shintani:2019wai, Davies:2019efs, Gerardin:2019rua, Borsanyi:2020mff, Lehner:2020crt, Aubin:2022hgm}
using four different classes of fermion discretizations. 
Whereas most results have an uncertainty of about 2\%, the result marked 
by `BMW 20' from \cite{Borsanyi:2020mff} has reached sub-percent precision.
As visible from the figure, this result is in between the experimental
result and the SM prediction from the R-ratio. A further, independent
lattice QCD result, preferably using a different fermion action, is needed
to make a clear statement concerning the current discrepancy between two 
types of SM predictions.

\begin{figure}
	\centering
	\includegraphics[width=.45\textwidth]{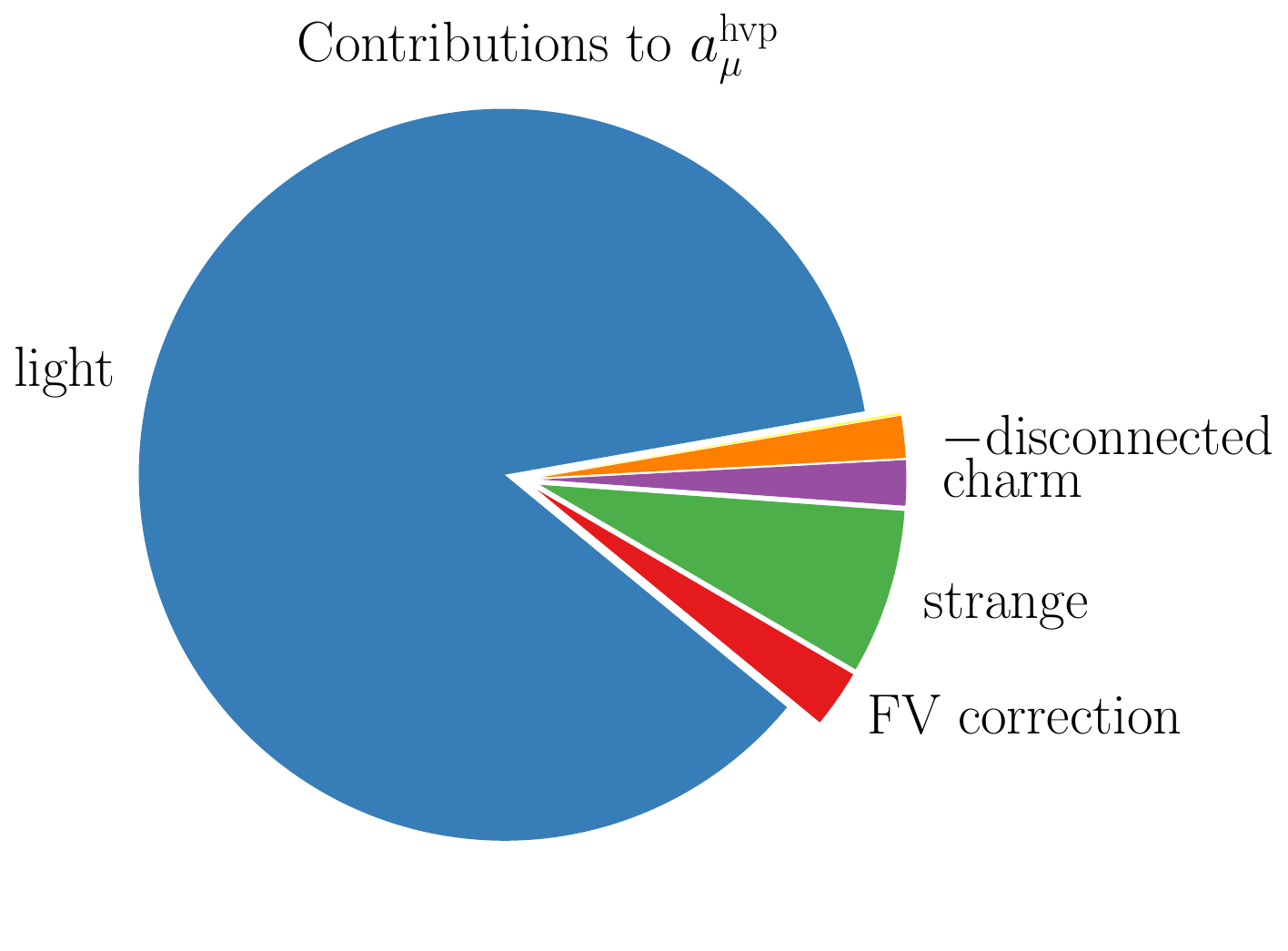}%
	\hspace{.1\textwidth}
	\includegraphics[width=.42\textwidth]{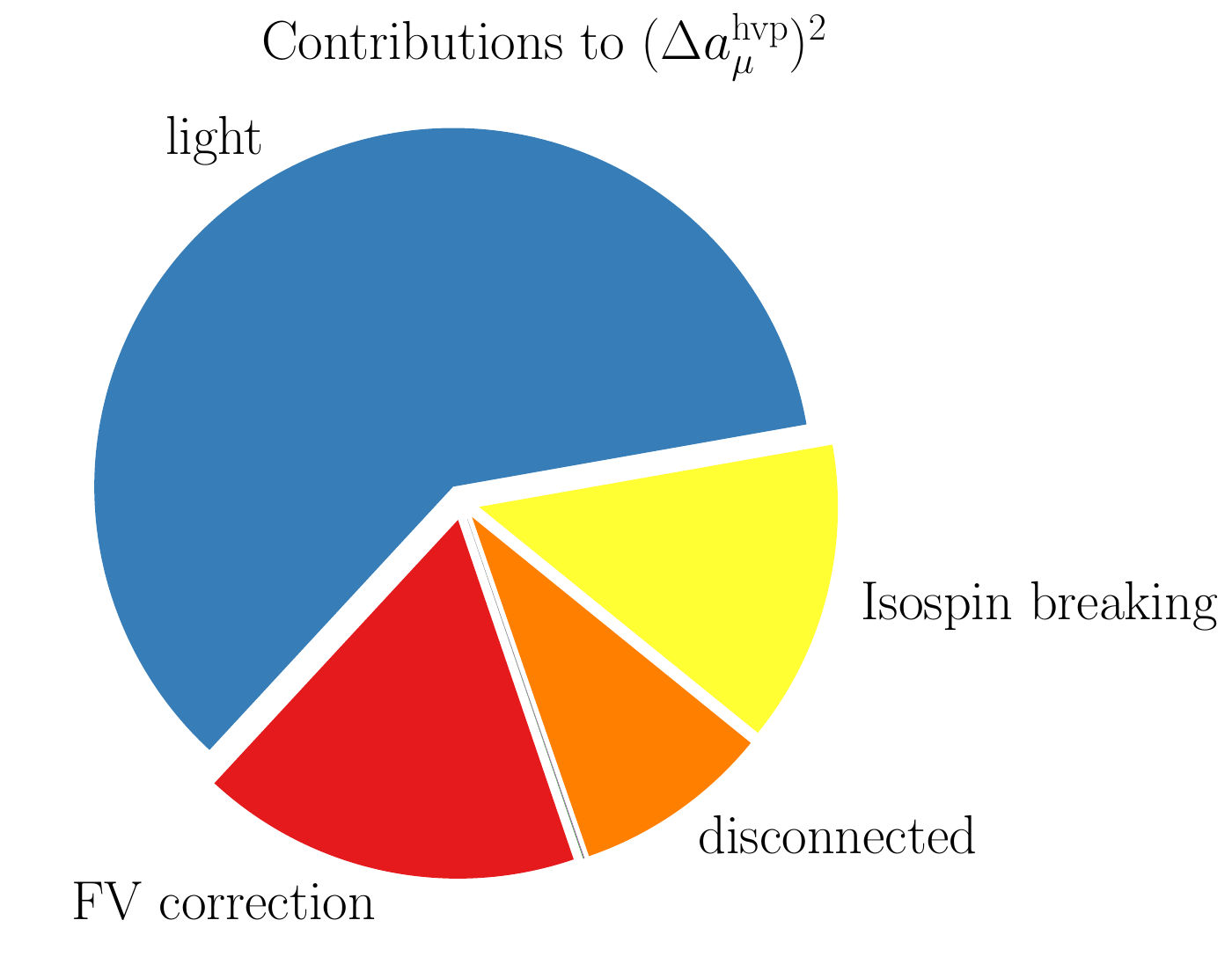}%
	\caption{ \label{f:contribs_amu}
		Contributions to the value (left) and the squared uncertainty (right) of $\ahvp$ based on the calculation in \cite{Borsanyi:2020mff}. The quark-disconnected contribution is negative.
	}
\end{figure}

The computation of $\ahvp$ on the lattice to sub-percent precision is
challenging due to a number of sources of uncertainty that have to 
be addressed appropriately in order to claim full control over the final
uncertainty. In Figure~\ref{f:contribs_amu} we illustrate the weight of
each contribution to $\amu$ and its uncertainty, based
on the result from \cite{Borsanyi:2020mff}.
On the left hand side, the contribution to the central value is shown, 
based on the decomposition in eq.~(\ref{e:Gt_flavdecomp}). 
About 90\% of the final value is due to the light-connected contribution.
Quark flavors with larger mass contribute less because the correlation 
function decays more rapidly. The quark-disconnected contribution is negative
and has a size of about 2\% of the total. As discussed above, finite-volume
effects in the flavor-decomposition mostly affect the light-connected and
disconnected contributions. 

On the right hand side of the figure, the contributions to the squared 
uncertainty of $\ahvp$ are shown. Since the result of \cite{Borsanyi:2020mff}
has reached sub-percent precision already, the pie chart gives an indication
of the sources of uncertainty that have to be addressed in order to further
improve the precision. Again, the dominant contribution comes from the 
light-connected contribution, containing the statistical uncertainty as 
well as the  systematic uncertainty from the continuum extrapolation.
The other three major contributions to the uncertainty are the correction
for finite-volume effects, the effects due to isospin breaking 
and the quark-disconnected contribution.

\section{Dominant sources of uncertainty}
In the following, we will describe the methods that are currently being used
to reduce and control the major sources of uncertainty in the calculation
of $\ahvp$.

\subsection{Controlling the long-distance tail}
The attainable statistical precision in the light-connected
(or, equivalently, isovector) contribution has been the dominant source of 
uncertainty of many lattice QCD calculations of $\ahvp$.
A dramatic improvement in precision is achieved by combining
theoretical constraints on the shape of the long-distance tail of 
the isovector correlation function with the use of
optimized estimators in its computation.

\begin{figure}
	\centering
	\includegraphics[width=.50\textwidth]{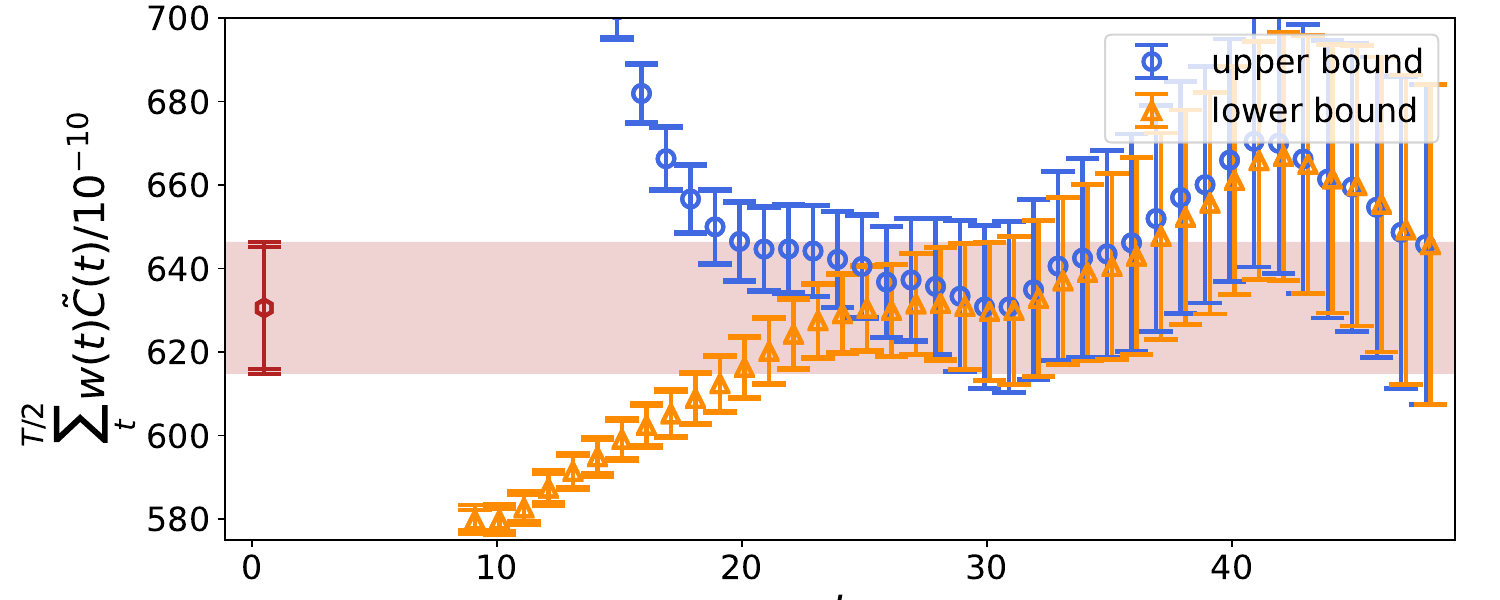}%
	\caption{\label{f:bounding}
		Bounding method applied to the local vector current correlation function. Figure taken from \cite{Bruno:2019nzm}.
	}
\end{figure}

Writing the spectral decomposition of the two-point correlation function
as
\begin{align}\label{e:Gt_spectral}
	G^{\rm I1}(t) = \sum_{n=0}^{N} \frac{Z_n}{2E_n}\mathrm{e}^{ -E_n t} + \mathrm{O}(\mathrm{e}^{- E_{N+1}t})\,,
\end{align}
with the (positive) overlap factors $Z_n$ and the finite-volume energies $E_n$, 
lower and upper bounds on the correlation function at Euclidean time $t$ 
are given by \cite{LehnerBounding2016, Budapest-Marseille-Wuppertal:2017okr, Blum:2018mom}
\begin{align}
	0 \leq G^{\rm I1}(t) \leq G^{\rm I1}(t_c) \mathrm{e}^{-E_0 (t - t_c)}\,,\qquad t\geq t_c\,.
\end{align}
The lower bound is based on the positivity of the correlation function and
may be further tightened, making use of the fact that the correlation function
in the long-distance regime decays more slowly than at short distances $\ll t_c$
\cite{Blum:2018mom, Gerardin:2019rua}.
The upper bound relies on an estimate of the ground state energy in finite 
volume that, when not known from data, has to be estimated conservatively
enough such that no model dependence is introduced when applying the bounds. 
As soon as both bounds agree with each other, the integration to infinite
distances in eq.~(\ref{e:def_TMR}) may be performed using either of the bounds
instead of the data itself. 
A example for this procedure is shown in Figure~\ref{f:bounding},
taken from \cite{Bruno:2019nzm}. 
A reduction of the error by a factor of 2 is reported by the authors.

\begin{figure}
	\centering
	\def \figwi {.2\textheight}
	\includegraphics[height=\figwi]{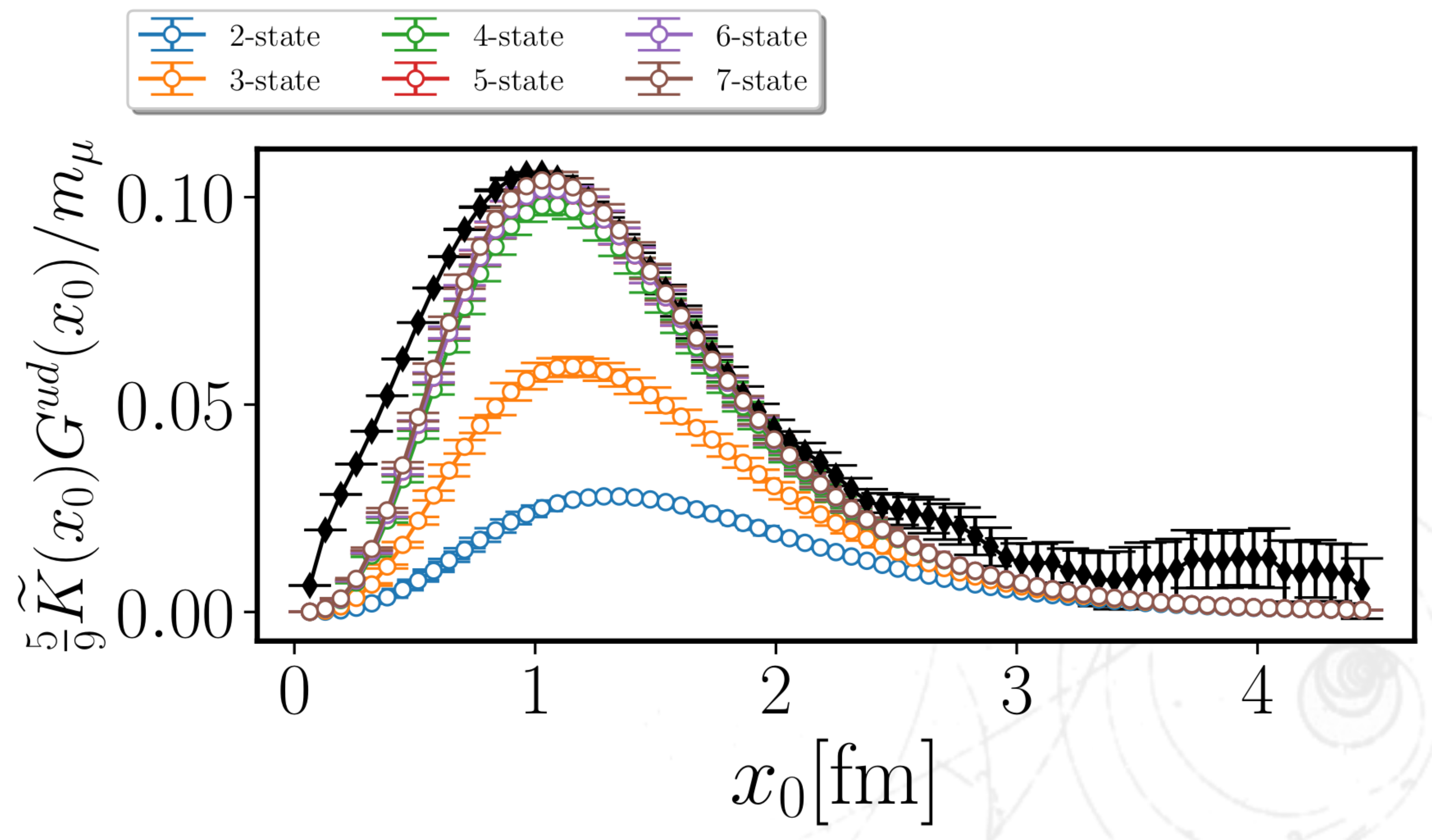}%
	\hfill
	\includegraphics[height=\figwi]{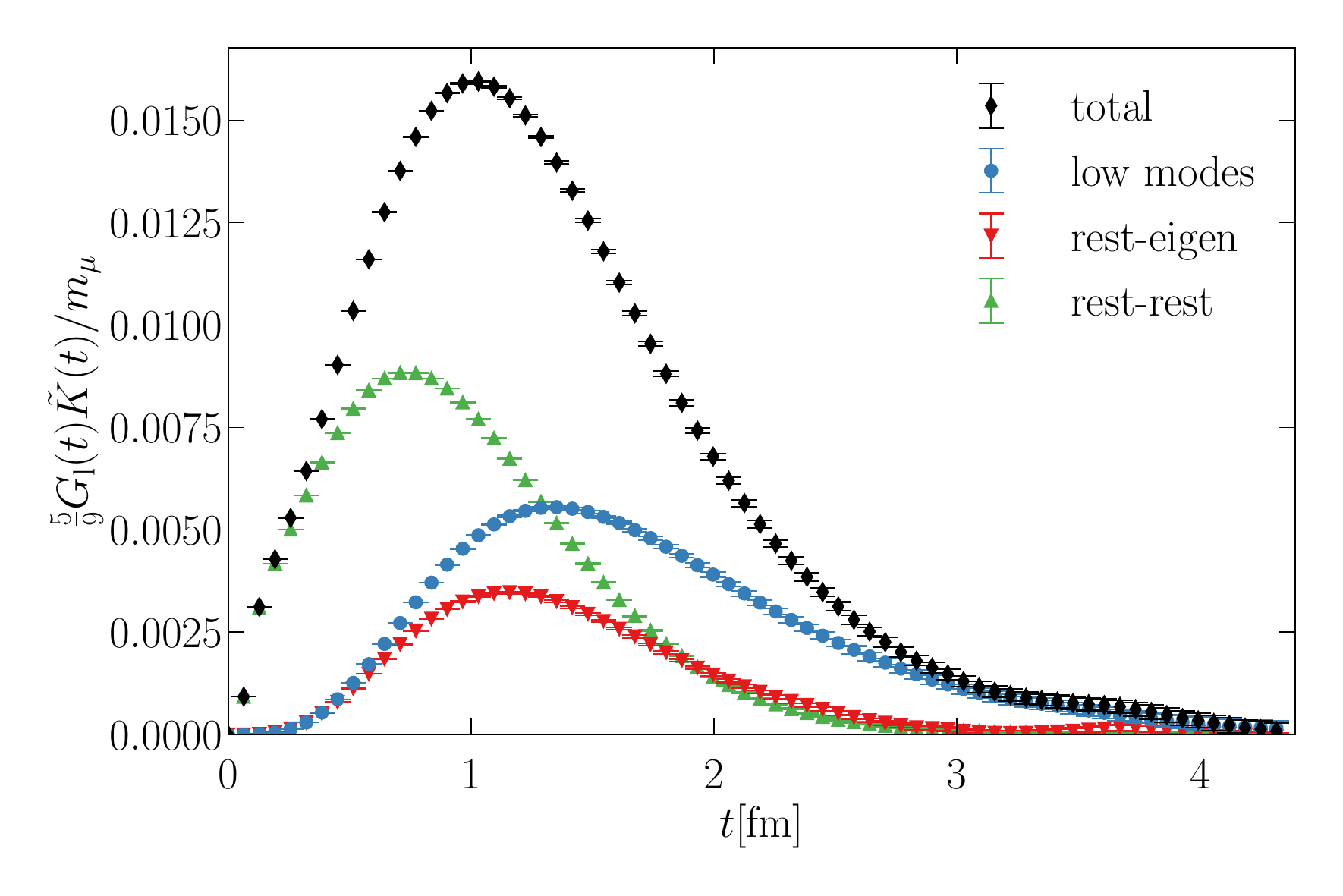}%
	\caption{\label{f:LMA} \label{f:spec}
		Left: Reconstruction of the integrand of the isovector contribution based on $n$ $\pi\pi$ states. The black diamonds show the corresponding integrand from stochastic sources, as also shown in Figure~\ref{f:TMR_integrand}. Taken from \cite{Paul:2023ksa}.
		Right: The same (blinded) integrand but computed using low-mode averaging. The black diamonds are the sum of the three colored contributions and the blue circles show the all-to-all evaluation based on 800 low modes of the Dirac operator.
	}
\end{figure}

A further significant reduction of uncertainty can be achieved if several 
of the lowest states contributing to the spectral decomposition in
eq.~(\ref{e:Gt_spectral}) are known. The corresponding finite-volume overlap
factors and energies may be computed in a dedicated spectroscopy analysis in
the isovector channel, as done in 
\cite{Gerardin:2019rua, Andersen:2018mau, Bruno:2019nzm}
The states coupling to the isovector current are mostly $\pi\pi$ states. 
In \cite{Bruno:2019nzm} it was found that $4\pi$ states seem to have a
negligible overlap with the current, even though their energies are among 
the smallest in the spectral decomposition.
Reconstructing the correlation function from the lowest-lying states eliminates
the signal-to-noise problem completely, since the error of the reconstructed correlation
function grows only linearly, as opposed to exponentially.
This is illustrated in the left-hand panel of Figure~\ref{f:spec}, taken 
from \cite{Paul:2023ksa}, which shows
that four $\pi\pi$ states are sufficient to saturate the isovector correlation
starting at $t \gtrsim 1.5\,$fm in a $(6.1\,{\rm fm})^3$ spatial volume 
at physical pion mass.

The use of advanced numerical procedures for the optimization of 
the estimator for $G^{\rm I1}(t)$ is common to all recent results for $\ahvp$.
Using the lowest $\mathrm{O}(1000)$ eigenmodes of the hermitian 
Dirac operator, at physical pion mass and $m_\pi L \approx 4$, to 
construct an all-to-all estimator of the long-distance tail of 
$G^{\rm I1}(t)$ 
\cite{Neff:2001zr, Giusti:2004yp, DeGrand:2004qw}
turns out to be significantly more efficient than sampling
with stochastic or point sources. 
Figure~\ref{f:LMA} shows an application of LMA for the same integrand
as the one shown in Figure~\ref{f:TMR_integrand}. It is
constructed by summing three contributions: The blue 
circles, labeled `low modes' show an all-to-all evaluation of the correlation
function in the subspace of the lowest 800 modes of the even-odd
preconditioned hermitian Dirac-Wilson operator. 
One clearly sees that this contribution to the integrand 
dominates starting at a distance of about $1.5\,$fm, exactly where the 
signal-to-noise problem starts to hinder the precise estimation of the
integrand. Whereas the all-to-all sampling of the correlation function in the
space of these lowest eigenmodes does not solve the exponential deterioration
of the signal, it reduces the coefficient of the growth significantly.
The correlation function in the orthogonal complement of the low modes,
indicated by green triangles in Figure~\ref{f:LMA} can be 
sampled stochastically. A number of methods exist to calculate the correlation
function that contains propagators from both subspaces which may be small
but has to be computed precisely to obtain a clean signal at large times.
The solution of the Dirac equation can be accelerated by inexact inversions
followed by a bias correction step 
\cite{Bali:2009hu, Blum:2012uh, Shintani:2014vja}.

The combination of some or all of the above techniques has enabled the community
to compute the isovector contribution to $\ahvp$, the largest source
of statistical uncertainty, with a precision of a few per-mil at physical
pion mass. 
We conclude this section by pointing out that new simulation
paradigms such as multi-level integration \cite{Ce:2016idq, Ce:2016ajy}
offer promising solutions to the signal-to-noise problem, 
as demonstrated in \cite{DallaBrida:2020cik} but not yet applied at large
scales.

\subsection{Finite-volume effects \label{s:fve}}
\begin{figure}
	\centering
	\includegraphics[height=.25\textheight]{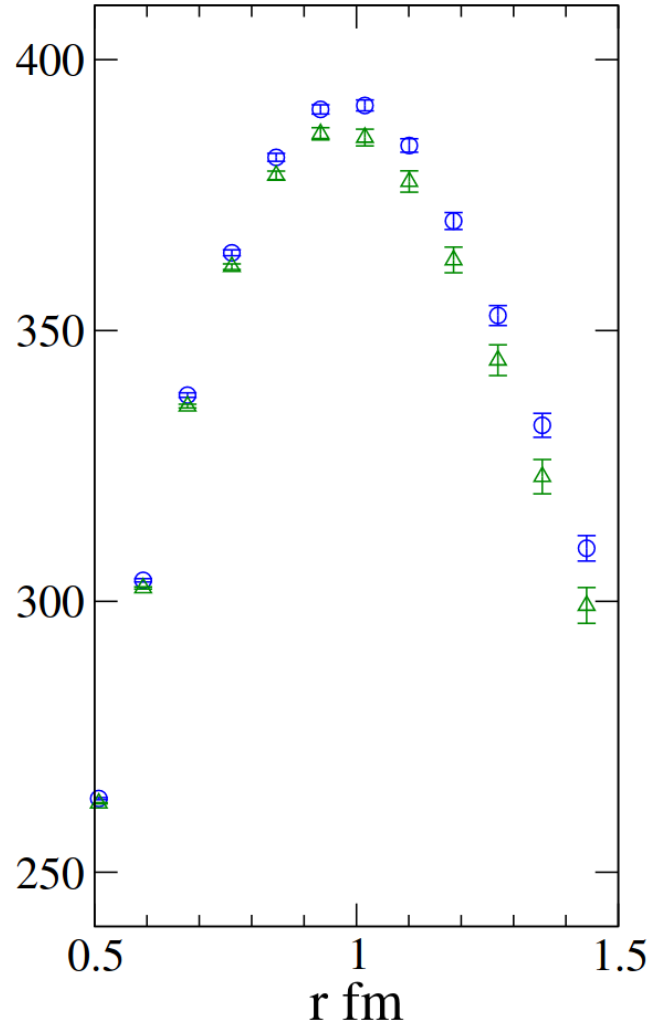}%
	\hspace{10mm}
	\includegraphics[width=.5\textwidth]{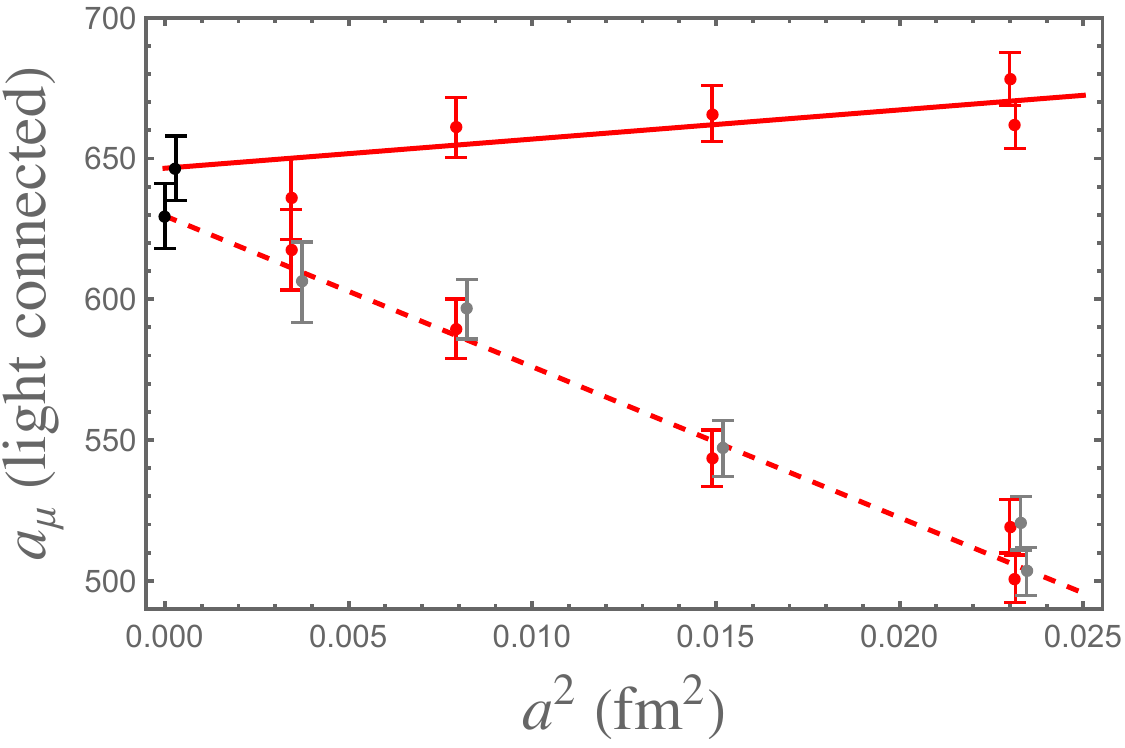}%
	\caption{\label{f:fv}\label{f:taste}
	Left: Zoom into the integrands of the light-connected contribution on lattices with spatial volumes (5.4\,fm)$^3$ (triangles) and (10.8\,fm)$^3$ (circles). Taken from \cite{Shintani:2019wai}.
	Right: Continuum limit of the light-connected contribution to $\ahvp$ with staggered quarks. The dashed line shows the continuum extrapolation without and the solid line the extrapolation with taste breaking corrections. Taken from \cite{Aubin:2022hgm}.
	}
\end{figure}

The numerically largest contributions to the finite-volume effects of $\ahvp$
enter in the long-distance regime of the light-connected channel which is
dominated by $\pi\pi$ states.
A significant effect is also present for the quark-disconnected contribution,
see section \ref{s:disc}, and isospin breaking effects, cf.\ section \ref{s:ib}.
At the physical value of the pion mass and a conventional value of $m_\pi L=4$, 
the finite volume shifts $\ahvp$ by about $3\%$ compared to its 
infinite-volume counterpart.
Any calculation aiming for sub-percent or even few per-mil precision 
therefore has to control the correction for finite-size effects at the 
$10\%$ level or use volumes which are significantly larger than
$(6\,{\rm fm})^4$. 

Several approaches have been worked out to perform a correction for 
finite-size effects based on effective field theories. It has been found that
next-to-leading (NLO) computations in chiral perturbation theory ($\chi$PT) 
can not fully describe finite-volume effects that are found in the data 
\cite{Shintani:2019wai, Giusti:2018mdh} 
and that NNLO effects are significant 
\cite{Aubin:2019usy, Bijnens:2017esv, Aubin:2020scy}. 
An extension of the framework incorporates the $\rho$ meson and photons
\cite{Sakurai:1960ju, Jegerlehner:2011ti, Chakraborty:2016mwy}.

The relation between the two-pion spectrum in finite volume and the 
timelike pion form factor can be used to compute the difference between
finite and infinite-volume isovector correlation functions based on a 
spectral decomposition \cite{Luscher:1986pf, Lellouch:2000pv, Meyer:2011um}.
This ansatz is expected to be particularly successful at large distances
where only a few states contribute to the finite-volume correlation function.
If no direct lattice calculation of the timelike pion form factor on the
lattice is available, a Gounaris-Sakurai parametrization \cite{Gounaris:1968mw} 
of the form factor can be employed for the finite-volume correction 
\cite{Francis:2013fzp, DellaMorte:2017dyu, Gerardin:2019rua, Giusti:2018mdh}.

A generic, relativistic effective field theory of pions has been employed to
compute finite-size and thermal effects on $\ahvp$ in 
\cite{Hansen:2019rbh, Hansen:2020whp}. It was suggested to approximate the
forward Compton amplitude by the pion pole term which is determined by the
electromagnetic pion form factor in the spacelike region.

Simulations in large volumes with spatial extents $L > 10\,$fm
\cite{Shintani:2019wai, Borsanyi:2020mff}, 
followed by a small, model-dependent, correction to infinite volume 
can be used to minimize the model dependence of finite-volume corrections.
Figure~\ref{f:fv} shows the integrand of the light-connected contribution to
$\ahvp$ for two spatial volumes of (5.4\,fm)$^3$ and (10.8\,fm)$^3$ from 
\cite{Shintani:2019wai}. A clear difference between the two can be resolved,
already in the intermediate-distance region. 
While simulations with $L>10\,$fm are prohibitively expensive at fine 
lattice spacing, they may be carried out for coarser lattices, 
together with an assumption on the size of the cutoff effects for the 
finite-volume correction. 
This approach has been used by the BMW collaboration
in \cite{Borsanyi:2020mff} to correct from a reference volume with 
$L\approx 6.3\,{\rm fm}$ and $T = \textstyle\frac{3}{2}L$ to a volume of $(10.75\,{\rm fm})^4$. The latter has been simulated at a single lattice 
spacing using an action with reduced taste-breaking effects.

\subsection{The continuum limit \label{s:cont}}
The systematic uncertainty associated with the continuum extrapolation can easily be the dominant source of uncertainty, as it is the case in the most
precise calculation to date \cite{Borsanyi:2020mff}. Whereas all of
the lattice results in Figure~\ref{f:comp_amu} employ a setup where cutoff
effects of $\mathrm{O}(a)$ are absent, higher-order cutoff effects and
modifications of the leading $a^2$ behavior may induce significant systematic
uncertainties since they may be difficult to constrain. The use of
at least four values of the lattice spacing and a sufficiently large range 
of resolutions is mandatory to claim control over the continuum extrapolation
in high-precision calculations of $\ahvp$.

As pointed out in \cite{Ce:2021xgd, Sommer:2022wac}, cutoff effects of 
$\mathrm{O}(a^2 \log(a))$ can be expected already at the classical level
of Symanzik effective theory (SymEFT) from the integration at very short
distances in the TMR integral in eq.~(\ref{e:def_TMR}) and the interplay
between correlation function and kernel function $\widetilde{K}(t)$.
These may be efficiently eliminated from the computation by using 
perturbative QCD at these short distances, as already suggested 
in \cite{Bernecker:2011gh}.

More worrisome in terms of potentially large corrections to $a^2$ scaling 
that might not be resolved at present-day lattice spacings 
are logarithmic modifications of the form 
$a^2 \rightarrow a^2[\bar{g}^2 (1/a)]^{\hat{\gamma}_i}$ with the renormalized
coupling $\bar{g}^2 (1/a)$ and one-loop anomalous dimensions $\hat{\gamma}_i$.
These effects are induced by quantum corrections in SymEFT. The dominantly
contributing anomalous dimensions from Wilson-type and Domain Wall 
fermion actions have been computed in \cite{Husung:2019ytz, Husung:2021mfl}
with the result that negative powers, which would slow down the approach to the
continuum limit, have not been found. 
An extension to quark bilinears has been presented at this
conference~\cite{Husung:Lattice2023}. 
In view of the target precision it is mandatory to include this 
information in continuum extrapolations of precision observables, such as
$\ahvp$, or to test a generous range of anomalous dimensions if no information
is available, as it is the case for staggered fermions. 

For staggered quarks, another significant source of discretization effects
is present in calculations of $\ahvp$, namely taste-breaking effects that
distort the pion spectrum at finite lattice spacing.
Since non-linearities in the continuum extrapolation are introduced by
the taste-breaking effects \cite{Aubin:2022hgm}, a correction is conventionally 
applied prior to the continuum extrapolation, based on staggered $\chi$PT
\cite{Davies:2019efs, Aubin:2019usy, Borsanyi:2020mff, Aubin:2022hgm}.
On the right hand side of Figure~\ref{f:taste} we show an example from 
a recent calculation \cite{Aubin:2022hgm} 
where the data is shown prior to the correction (lower data points) 
and after the correction. 
Care has to be taken in this step to ensure that all possible model dependence 
is appropriately accounted for in the estimation of the systematic uncertainty.

\begin{figure}
	\centering
	\def \figwi {.46\textwidth}
	\includegraphics[width=\figwi]{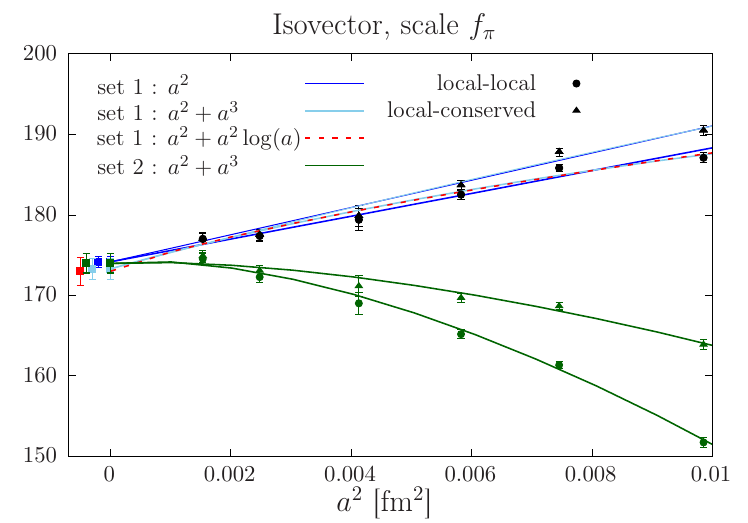}
	\hspace{.05\textwidth}
	\includegraphics[width=.41\textwidth]{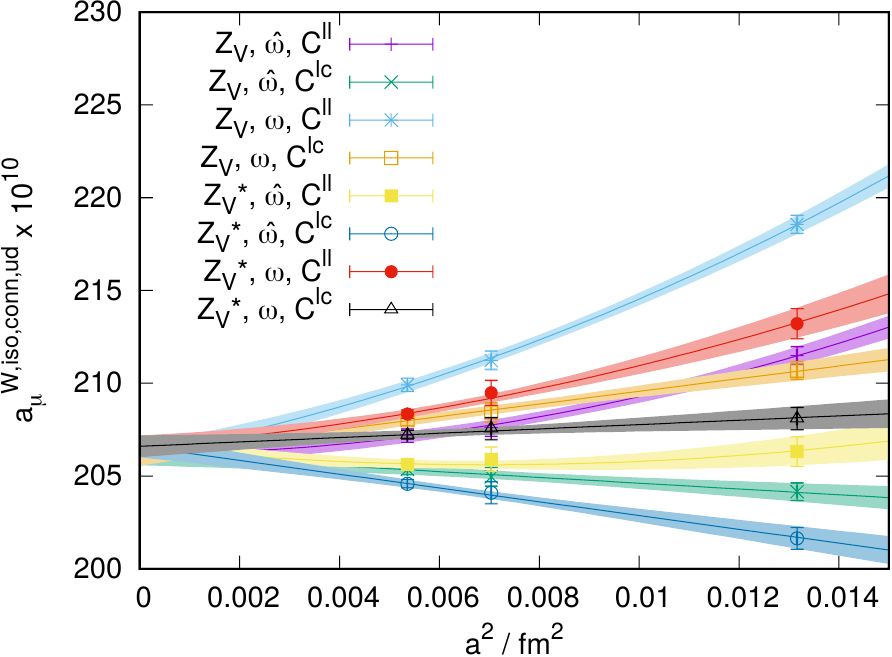}
	\caption{\label{f:cont_id}
		Continuum extrapolations of the isovector contribution to $\awin$. Each data set and the corresponding extrapolation is based on a specific choice to define the observable at finite lattice spacing.
		Left: At the SU(3) symmetric point from \cite{Ce:2022kxy}.
		Right: At physical quark mass from \cite{RBC:2023pvn}. 
	}
\end{figure}

Various definitions of the observable at finite lattice spacing, differing 
by $\mathrm{O}(a^2)$, have been applied when computing window observables. 
This approach serves a dual purpose: either as a cross-check ensuring that
extrapolations coincide in the continuum limit \cite{Ce:2022kxy} 
or through combined extrapolations that are constrained to be consistent 
in the continuum \cite{ExtendedTwistedMass:2022jpw, RBC:2023pvn}.
Two of these extrapolations for the isovector contribution to the
intermediate-distance window observable are shown in Figure~\ref{f:cont_id}.
While this approach constrains cutoff effects in the valence sector, cutoff
effects from the sea that are common to all data points are not better
constrained.

\subsection{Scale setting \label{s:scale}}
Although $a_\mu$ is a dimensionless quantity it inherits a scale
dependence from the conversion of the muon mass to lattice units
in the QED kernel function $\widetilde{K}$. Denoting the scale
setting quantity by $\Lambda$, we can write
\begin{align}\label{e:scaledep_ahvp}
	\frac{\partial\ahvp}{\partial\Lambda}= \left(\frac{\alpha}{\pi}\right)^2\int_0^{\infty} dt\,G(t)
	\left(\frac{\partial}{\partial\Lambda}\widetilde{K}(t)\right)\,,
\end{align}
and, as pointed out in \cite{DellaMorte:2017dyu}, a relative uncertainty
$\Delta \Lambda / \Lambda$ on the scale setting quantity leads to a contribution
of $\Delta \ahvp / \ahvp \approx 1.8\, (\Delta \Lambda / \Lambda)$ to the 
uncertainty of $\ahvp$.
Ultimately, a scale determination at the per-mil level has to be performed
to be able to reach the precision goal for $\ahvp$. 
The spread of results for (ratios of) gradient flow scales
reported in the recent FLAG report \cite{Aoki:2021kgd} is an indication
that the control over all sources of uncertainty may not be sufficient,
at present.
Aiming at a complete calculation of $\ahvp$ with sub-percent precision,
the effect of isospin breaking has to be included in the scale setting.
The $\Omega$ baryon, not being affected at leading order by the effect
of strong isospin breaking, has emerged as the preferred scale setting 
quantity for many collaborations. 
Here, the main difficulty is the reliable extraction of the ground state
mass from the baryon correlation function.

Incorporating a generic scale dependent window function $w^X(t)$ into 
eq.~(\ref{e:scaledep_ahvp}) leads to
\begin{align}\label{e:scaledep_awin}
\frac{\partial(\ahvp)^X}{\partial \Lambda}= \left(\frac{\alpha}{\pi}\right)^2\int_0^{\infty} dt\,G(t)
\left[
\left(\frac{\partial}{\partial \Lambda} w^X(t)\right) \widetilde{K}(t)
+
w^X(t) \left(\frac{\partial}{\partial \Lambda}\widetilde{K}(t)\right)
\right]\,.
\end{align}
An analysis of the interplay of the window functions of 
eq.~(\ref{e:def_windows}) with $\widetilde{K}$ shows that the
short-distance window has barely any scale dependence. For the 
intermediate-distance window observable one finds
$\Delta (\ahvp)^{\rm id} / (\ahvp)^{\rm id} \approx 0.5\, (\Delta \Lambda / \Lambda)$.
The significantly reduced scale dependence of $\awin$ further facilitates 
the precise computation of this quantity from lattice QCD.
Consequently, an enhanced scale dependence with respect to $\ahvp$ is found 
for the long-distance window observable.

We note that high-precision comparisons and consistency checks between lattice
QCD results can be performed if the same, appropriate scale setting quantity
is used throughout. Since the lattice uncertainty on an experimentally known
dimensionful quantity does not enter in this case, gradient flow scales seem to
be best suited for these comparisons as they can be computed with 
negligible uncertainty at finite lattice spacing 
(as opposed to baryon masses or decay constants).

\section{Euclidean time windows in the time-momentum representation \label{s:windows}}
The three window quantities defined in eq.~(\ref{e:def_windows}) 
are the ideal testbed to verify the control over the systematic effects,
which have been
outlined in the previous section, by comparison among lattice results.
Whereas finite-size effects contribute significantly in the long-distance
window, where also the loss of signal plays an important role, a computation 
of the short distance window observable will be mostly affected 
by the systematic uncertainty from taking the continuum limit.
The intermediate distance window observable on the other hand is designed such
that these systematic effects are suppressed and therefore allows to compute 
and compare lattice results with high precision. 

The windows in Euclidean time of eq.~(\ref{e:def_windows}) can be translated 
to momentum space, allowing for a calculation of the window
observables from the R-ratio \cite{Borsanyi:2020mff, Colangelo:2022vok}
and a comparison with lattice results.
Since the localized windows in Euclidean time are non-local in momentum
space, it is non-trivial to ascribe possible discrepancies between the
two approaches to a certain interval in momentum space. Different window 
observables than the ones of eq.~(\ref{e:def_windows}) have been suggested 
in \cite{Aubin:2022hgm, Colangelo:2022vok, Boito:2022njs} 
to allow for a more detailed comparison with dispersive methods .
The combination of several observables can also help to disentangle
contributions from different energy regions, as outlined in
\cite{Davier:2023cyp, Colangelo:2022xfy}.

A complementary path for a direct comparison between lattice results and
the experimentally determined R-ratio has been taken in 
\cite{ExtendedTwistedMassCollaborationETMC:2022sta}
where the smeared R-ratio has been obtained from the lattice using 
reconstruction methods and compared to its experimental counter part.
The authors report a three sigma tension in the $\rho$ region with respect
to the R-ratio evaluation of \cite{Keshavarzi:2019abf}.

\subsection{The intermediate-distance window \label{s:id}}
\begin{figure}
	\centering
	\includegraphics[width=.49\textwidth]{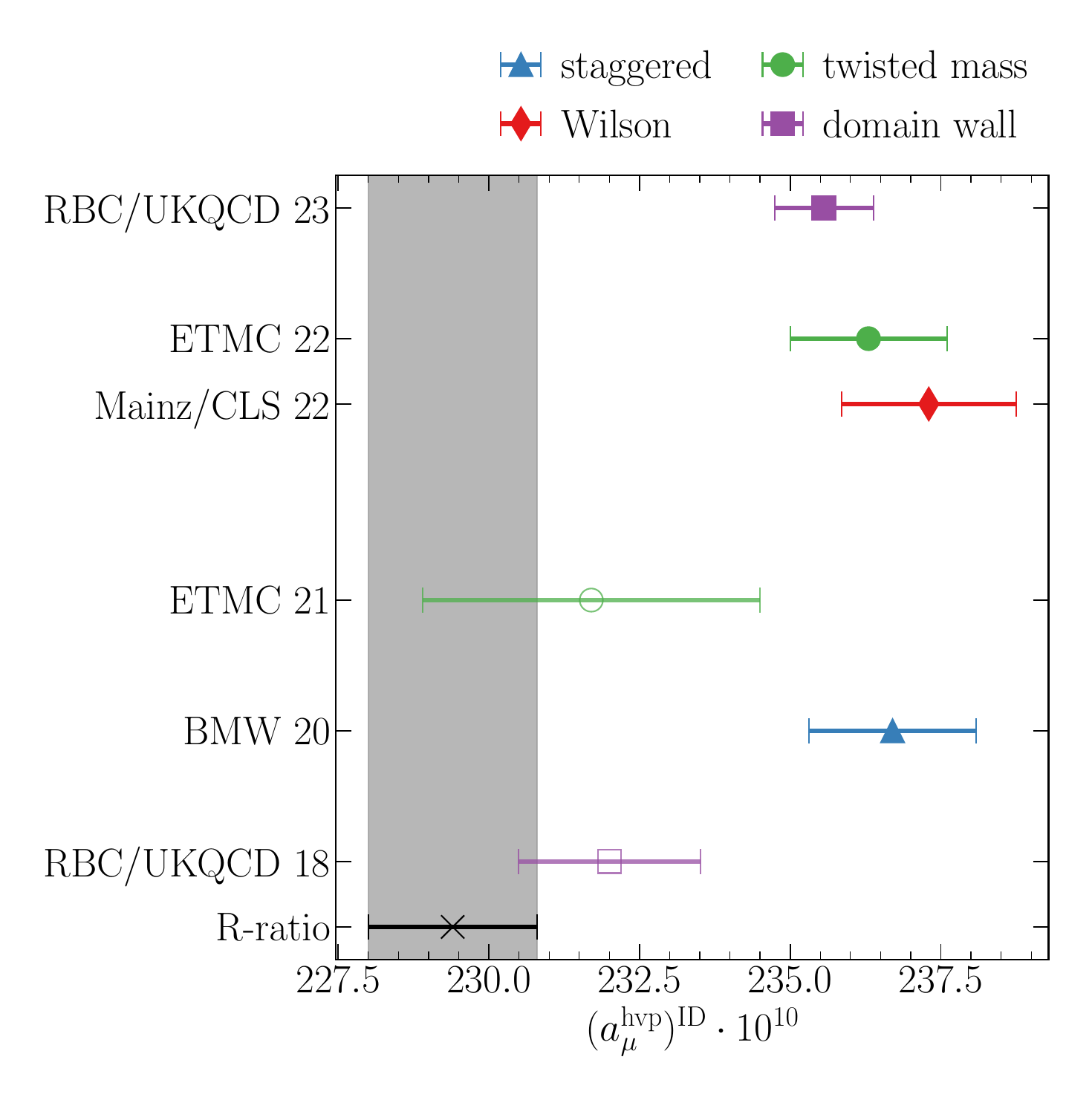}%
	\hfill
	\includegraphics[width=.49\textwidth]{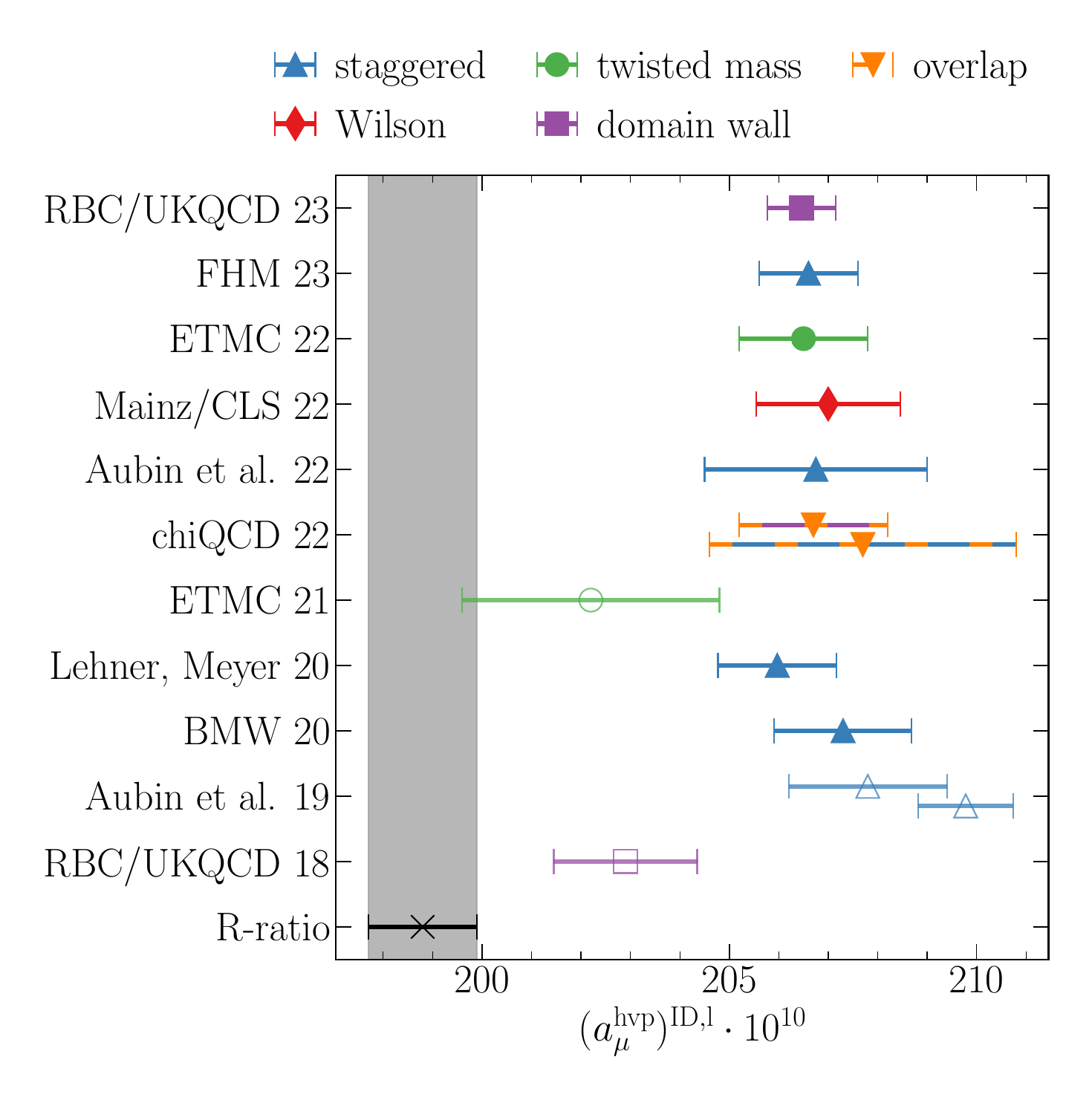}%
	\caption{\label{f:id}
	Left: Evaluations of $\awin$ from lattice QCD (colored points)
	\cite{Blum:2018mom, Borsanyi:2020mff, Giusti:2021dvd, Ce:2022kxy, ExtendedTwistedMass:2022jpw, RBC:2023pvn} and the R-ratio \cite{Colangelo:2022vok}.
	Right: Evaluations of $\awinl$ from lattice QCD (colored points)
	\cite{Blum:2018mom, Aubin:2019usy, Borsanyi:2020mff, Lehner:2020crt, Giusti:2021dvd, Wang:2022lkq, Aubin:2022hgm, Ce:2022kxy, ExtendedTwistedMass:2022jpw, FermilabLatticeHPQCD:2023jof, RBC:2023pvn} and the R-ratio \cite{Benton:2023dci}.
	chiQCD 22\,\cite{Wang:2022lkq} used overlap fermions in the valence sector and domain wall or staggered quarks in the sea.
	Superseded results are denoted by open symbols.
	}
\end{figure}

Significant progress has been made in the computation of the 
intermediate distance window observable in the past years. 
Four collaborations 
\cite{Borsanyi:2020mff, Ce:2022kxy, ExtendedTwistedMass:2022jpw, RBC:2023pvn} 
have performed a complete calculation of $\awin$
with uncertainties between 0.6\% and 0.35\% which are dominated by
systematic uncertainties in all cases.%
\footnote{The estimate for isospin breaking effects in the result of 
\cite{ExtendedTwistedMass:2022jpw} is taken from \cite{Borsanyi:2020mff}.}
The four results agree with each other and thus provide a confirmation
of the result of \cite{Borsanyi:2020mff} in the intermediate distance
region that makes up for about one third of $\ahvp$. 
The results are denoted by filled colored symbols on the left hand side of
Figure~\ref{f:id} together with an evaluation of the same quantity 
from the R-ratio \cite{Colangelo:2022vok}.%
\footnote{
The open symbols denote the results from \cite{Blum:2018mom, Giusti:2021dvd}
which are superseded by the recent results of the RBC/UKQCD and ETM 
collaborations. The differences with respect to their recent results
are understood to be due to long continuum or chiral extrapolations,
respectively.}

If the four lattice results are averaged, assuming
a very conservative 100\% correlation, a $3.8\,\sigma$ discrepancy with respect
to the dispersive result is found \cite{Wittig:2023pcl}.
This points towards underestimated systematic 
uncertainties in either of the two approaches. 
Since the finite-volume corrections for $\awin$ are smaller than the final
uncertainties and four different quark actions have been used, it is
challenging to find an explanation for a potential effect that would 
affect all lattice results similarly and lead to a shift of about $3.5\%$. 
We note that the inclusion of the recent result of the CMD-3 collaboration 
for the $e^+e^-\to \pi^+\pi^-$ cross section at low energies
in the data-driven evaluation of $\awin$ would lead to a larger value
that would be better compatible with lattice QCD \cite{CMD-3:2023alj}. 
A similar observation is made when $\tau$ spectral functions are
used instead of $e^+e^-$ results 
\cite{Miranda:2020wdg, Masjuan:2023qsp, Davier:2023fpl}.
The current $3.8\,\sigma$ discrepancy in $\awin$ 
is about half of the difference between the White Paper average and the
result of \cite{Borsanyi:2020mff} for $\ahvp$.

A recent result for the light-connected contribution to
the intermediate distance window observable, $\awinl$, 
from the R-ratio \cite{Benton:2023dci} can be compared to 
a larger number of lattice results for this dominant
contribution to $\awin$, as shown on the right-hand 
side of Figure~\ref{f:id}.
We note that only one further independent set of gauge ensemble is added 
when including the additional results. The two most recent results
are based on blinded analyses to prevent a human bias in view of the large
number of previous results.
The discrepancy between lattice and R-ratio is even larger for $\awinl$
due to smaller uncertainties and the fact that no significant deviation
is found in the sum of strange-connected and disconnected contributions
\cite{Benton:2023dci}. 
This can be taken as indication that an insufficient description 
of lattice artefacts cannot be responsible for the large discrepancy 
as it would affect light-connected and  strange-connected 
contributions in a similar manner.

All of the results shown in Figure~\ref{f:id} use the time-momentum 
representation to compute $\awinl$. A calculation based on the Lorentz-covariant
coordinate-space representation \cite{Meyer:2017hjv} has confirmed the 
result of \cite{Ce:2022kxy} at a pion mass of $350\,$MeV \cite{Chao:2022ycy}.
It is expected that this representation is especially well suited for the 
combination with simulations in the master-field paradigm 
\cite{Francis:2019muy}.

\subsection{The short-distance window \label{s:sd}}
The control of cutoff effects from small Euclidean distances is the main challenge in any computation of the short-distance window observable $\asd$.
Therefore, a successful comparison of lattice results can provide an 
indication that these cutoff effects are fully under control. 
From the phenomenological point of view, the comparison with an R-ratio
evaluation might give further information concerning the origin of the 
current discrepancies.%
\footnote{An analysis of a phenomenological model 
for the R-ratio \cite{Bernecker:2011gh}
suggests that, if the discrepancy in $\awin$ is due to an 
underestimate of the experimentally determined R-ratio in the region between
600\,MeV and 900\,MeV, a corresponding discrepancy of more than one percent 
would be expected in $\asd$ \cite{Ce:2022kxy}.
}

To date, one complete calculation \cite{ExtendedTwistedMass:2022jpw} 
and one calculation of the light-connected part to $\asd$ 
\cite{RBC:2023pvn} have been published.
Both works employ three values of the lattice spacing in their combined
continuum extrapolations of observables that have to agree in the
continuum limit. The subtraction of tree-level cutoff effects 
\cite{ExtendedTwistedMass:2022jpw} and the use of perturbation theory 
at order $\alpha_{\rm s}^4$  at very short distances \cite{RBC:2023pvn}
are used to tame or remove the potentially dangerous
$\mathrm{O}(a^2 \log(a))$ cutoff effects from the TMR integral.
Whereas the two lattice results for $\asdl$ agree,
\cite{ExtendedTwistedMass:2022jpw}
finds a slight $1.4\,\sigma$ or $1.3\%$ shift compared to the R-ratio evaluation
of \cite{Colangelo:2022vok}.

\section{Subleading contributions}
In addition to the dominant sources of uncertainty in lattice computations
of $\ahvp$, which primarily affect the light-connected contribution,
two further significant sources of uncertainty can be identified on the
right hand side of Figure~\ref{f:contribs_amu}, namely the one from the quark-disconnected contribution and from the inclusion of isospin breaking
effects.

\subsection{The quark-disconnected contribution \label{s:disc}}
Neglecting charm quark contributions that have been shown to be small
\cite{Borsanyi:2017zdw, ExtendedTwistedMass:2022jpw}, 
the quark-disconnected contribution can be 
formulated from differences of light and strange quark loops 
\cite{Gulpers:2014jaq}. 
Therefore, this contribution vanishes at the SU(3) symmetric point of 
QCD, where it is a double zero in the quark mass difference
$(m_{\rm l} - m_{\rm s})$ \cite{Gerardin:2019rua}.

The estimation of this contribution is complicated by the fact that
the noise of the quark-disconnected correlation function is constant 
whereas the correlation function falls off exponentially. At very long
distances, it can be expected that the ratio of quark-disconnected and 
isovector correlation function approaches the asymptotic ratio
$-\textstyle\frac{1}{9}$ \cite{DellaMorte:2010aq, Francis:2013fzp}.
Hence, the quark-disconnected contribution is foreseen to exhibit significant
finite-volume effects, approximately $\textstyle\frac{1}{9}$ those 
observed in the isovector channel.
A cancellation of these effects is obtained in the isoscalar contribution,
see eq.~(\ref{e:Gt_isodecomp}).

All-to-all propagators have to be computed in order to calculate 
the noisy quark-disconnected contribution. Several advanced algorithmic 
techniques are applied in this calculation, in addition to stochastic
sources. As for the isovector correlation function,
low modes of the Dirac operator and truncated solves \cite{Bali:2009hu} are 
used by many collaborations. Hierarchical probing \cite{Stathopoulos:2013aci},
randomized sparse grids \cite{Blum:2015you} or the hopping parameter 
expansion \cite{Thron:1997iy} 
for heavy quark contributions may lead to further improvements.
Exploiting the structure of the correlation function by using the same 
set of sources for light and strange quark inversions has a significant
impact on the quality of the signal \cite{Gulpers:2014jaq}. 
The cancellation of noise in differences is further exploited in
frequency-splitting estimators \cite{McNeile:2006bz, Giusti:2019kff}.
A variant of the bounding method for the quark-disconnected
\cite{Borsanyi:2017zdw} or the isoscalar contribution \cite{Ce:2022eix} 
can be used to enhance the precision of the TMR integral.

The precision of published evaluations of the disconnected contribution to
$\ahvp$ is approaching the level of $10\%$
\cite{Blum:2018mom, Davies:2019efs, Gerardin:2019rua, Borsanyi:2020mff}, 
not far from the target that is needed to reach per-mil precision for $\ahvp$.

\subsection{Isospin breaking effects \label{s:ib}}
As soon as the precision of a Standard Model prediction has reached 
the $1\%$ level, it is not sufficient to work in isospin symmetric QCD.
Effects from strong-isospin breaking of order 
$(m_{\rm d} - m_{\rm u})/ \Lambda_{\rm QCD}$ and from the inclusion of
QED at order $\alpha$, have to be considered, see, e.g., 
\cite{Patella:2017fgk, DiCarlo:Lattice2023}.

Whereas simulations with dynamical QCD+QED 
\cite{Westin:2020blm, RCstar:2022yjz} 
and non-degenerate up and down quarks \cite{FermilabLattice:2017wgj} 
are pursued in the context of computations of $\ahvp$,
most calculations of isospin breaking effects in
$\ahvp$ are performed via a perturbative expansion in the
isospin breaking parameters $\Delta m_f = m_f - \bar{m}$ 
and $\alpha$ around isospin symmetric QCD on isospin symmetric 
gauge ensembles \cite{deDivitiis:2011eh, deDivitiis:2013xla}. 
The benefit of being able to reuse existing gauge fields comes at the 
cost of having to compute a number of additional correlation functions.
The combination of infinite-volume QED with finite-volume QCD has been
proposed as alternative 
\cite{Blum:2018mom, Biloshytskyi:2022ets, Chao:2023lxw}, 
similar to calculations of $\ahlbl$ \cite{Green:2015sra, Asmussen:2016lse}.

When QED$_{\rm L}$ \cite{Hayakawa:2008an} is used to formulate QED inside 
a finite box with periodic boundary conditions in 
the spatial directions, as it is the case in many calculations, 
the finite volume effects scale with $1/L^3$ \cite{Bijnens:2019ejw},
where $L$ is the spatial extent.
Given the conventionally used lattice sizes, this
is sufficient in view of the precision targets since the finite-volume
corrections affect a small contribution to $\ahvp$. We note that the 
corrections are reduced to order $1/L^4$
when using QED$_{\rm C}$ \cite{Polley:1990tf, Lucini:2015hfa}
or QED$_{\rm r}$ \cite{DiCarlo:Lattice2023}.

When comparing results for the isospin breaking effects in $\ahvp$ or
results for $\ahvp$ in isopin symmetric QCD, care has to be taken
since the separation of strong and QED isospin breaking effects is
scheme dependent, as is the definition of the physical point in isosymmetric
QCD. Whereas the differences between commonly used schemes are expected to
be small, a common scheme prescription for the lattice community would 
significantly facilitate such comparisons in view of shrinking uncertainties
\cite{Tantalo:2023onv}.
The definition of the physical point in full QCD+QED is unambiguous.

No new results for isospin breaking effects in $\ahvp$ have been published since the review in \cite{Gerardin:2020gpp}. The effect of strong isospin
breaking on the quark-connected contribution to $\ahvp$  has been computed in
\cite{FermilabLattice:2017wgj, Blum:2018mom, Giusti:2019xct, Borsanyi:2020mff, Lehner:2020crt}. 
The BMW collaboration found a large cancellation with the corresponding effect
in the quark-disconnected contribution \cite{Borsanyi:2020mff}. 
QED effects in the valence sector of quark-connected 
\cite{Blum:2018mom, Giusti:2019xct, Borsanyi:2020mff} and quark-disconnected
\cite{Blum:2018mom, Borsanyi:2020mff} have been computed by several 
collaborations, however, partly with 100\% uncertainties. 
Only the calculation in \cite{Borsanyi:2020mff} includes the effects of
isospin breaking on sea quarks by computing diagrams that are suppressed
by SU(3)$_f$ or $1/N_c$ on boxes with spatial sizes of about $3\,$fm.

The complete calculation of \cite{Borsanyi:2020mff} finds significant
cancellations between different contributions to isospin breaking
effects. Their final result of $0.5(1.4)\cdot 10^{-10}$ amounts to a
sub per-mil effect on $\ahvp$. In view of the difficulty of obtaining precise
results for the isospin breaking effects, the smallness of the contribution
is encouraging. However, more independent and precise results for the QED 
effects on valence and sea quarks are required before definite conclusions
can be drawn. Progress in this direction has been made by several
collaborations in the last years, see e.g.
\cite{Ce:2022kxy, Ray:2022ycg, Harris:2023zsl}.

\section{Conclusions}
The upcoming decrease of experimental uncertainties for the anomalous
magnetic moment of the muon after the analysis of the Run-4-6 data of
the Fermilab Muon $g-2$ experiment calls for a reduction of theory
uncertainties, especially for the HVP contribution.
However, the disagreement between the data-driven results 
that were included in the White Paper average \cite{Aoyama:2020ynm} and
the most precise result from lattice QCD \cite{Borsanyi:2020mff}
has posed new questions concerning the SM prediction. 
A recent result \cite{CMD-3:2023alj} for the $e^+e^-\to \pi^+\pi^-$
cross-section at low energies that is incompatible with the two 
most precise previous results calls into question the control of 
systematic uncertainties of the data-driven calculation.

Lattice QCD computations have reached the stage where they are able
to provide precise and 
reliable SM predictions for hadronic contributions to $a_\mu$.
Several independent results for $\ahvp$ with sub-percent precision 
are needed to confirm that all sources of uncertainty are under 
control. Thanks to algorithmic advances, a deeper theoretical
understanding of finite-volume and lattice spacing effects and 
a significant amount of invested computing resources, this goal is in reach.

For the intermediate distance window observable, the high degree
of agreement between several independent results, as shown in 
Figure~\ref{f:id}, leads to the conclusion that the discrepancy
with respect to data-driven evaluations is firmly established. 
Given various cross-checks, a common systematic effect that affects all
lattice results seems very unlikely. 

It is desirable to obtain a similar level of agreement of lattice results
for $\ahvp$. With the ongoing reduction of statistical uncertainties, 
control of systematics becomes even more important. Calculations in 
very large boxes play an important role to reduce a possible 
model dependence of the finite-size correction. 
For upcoming results, the systematic 
uncertainty from the continuum extrapolation will likely become the
dominant source of uncertainty. Since the accessible range of lattice spacings
is limited, mostly due to critical slowing down towards the continuum 
limit \cite{Schaefer:2010hu}, further significant improvements in the 
control of cutoff effects are difficult to achieve with simulations based
on the Hybrid Monte Carlo algorithm. Better informed extrapolations based 
on SymEFT may help to reduce systematic uncertainties.
In the meantime, combined continuum extrapolations of data from 
different fermion discretizations, having significantly different lattice 
artefacts, could constrain the approach to the continuum limit and thus
lead to reduced systematic uncertainties. 

Existing results indicate that isospin breaking effects in $\ahvp$ 
are significantly smaller than $1\%$.
More independent results are needed to confirm this conclusion.
If a common scheme for QCD+QED is used in upcoming calculations,
averages of sub-quantities, such as the light-connected contribution to
$\ahvp$, could be used to improve the global precision of $\ahvp$ from 
lattice QCD. 

Given the various tensions, blinded analyses have become standard to 
reduce the human bias in the analysis. 
We expect several new results for $\ahvp$ and $\ald$ to be published in
the next year. The outcome will be decisive in scrutinizing current 
tensions between Standard Model predictions and between theory and 
experiment.

\subsection*{Acknowledgements}
I thank the organizers for the invitation to present a summary
of the lattice calculations of $\ahvp$,
Harvey Meyer and Hartmut Wittig for a careful reading of the manuscript and all members of the Mainz lattice group for the fruitful
collaboration in the last years.
Calculations by Mainz/CLS were performed on HPC platforms at Mainz, JSC Jülich
and HLRS Stuttgart. The support of the Gauss Centre for Supercomputing (GCS) and the John von Neumann-Institut für Computing (NIC) for projects CHMZ21 and CHMZ23 at JSC and project GCS-HQCD at HLRS is gratefully acknowledged.
This project has received funding from the European Union’s Horizon Europe research and innovation programme under the Marie Sk\l{}odowska-Curie grant agreement No 101106243.

{
	\bibliographystyle{jhep_collab}
	\bibliography{biblio}

\providecommand{\href}[2]{#2}\begingroup\raggedright\begin{thebibliography}{100}

\bibitem{Fan:2022eto}
X.~Fan, T.~G. Myers, B.~A.~D. Sukra and G.~Gabrielse, \emph{{Measurement of the
  Electron Magnetic Moment}},
  \href{https://doi.org/10.1103/PhysRevLett.130.071801}{\emph{Phys. Rev. Lett.}
  {\bfseries 130} (2023) 071801}
  [\href{https://arxiv.org/abs/2209.13084}{{\ttfamily 2209.13084}}].

\bibitem{Bennett:2006fi}
{\scshape Muon g-2} collaboration, G.~W. Bennett et~al., \emph{{Final Report of
  the Muon E821 Anomalous Magnetic Moment Measurement at BNL}},
  \href{https://doi.org/10.1103/PhysRevD.73.072003}{\emph{Phys. Rev. D}
  {\bfseries 73} (2006) 072003}
  [\href{https://arxiv.org/abs/hep-ex/0602035}{{\ttfamily hep-ex/0602035}}].

\bibitem{Muong-2:2021ojo}
{\scshape Muon g-2} collaboration, B.~Abi et~al., \emph{{Measurement of the
  Positive Muon Anomalous Magnetic Moment to 0.46 ppm}},
  \href{https://doi.org/10.1103/PhysRevLett.126.141801}{\emph{Phys. Rev. Lett.}
  {\bfseries 126} (2021) 141801}
  [\href{https://arxiv.org/abs/2104.03281}{{\ttfamily 2104.03281}}].

\bibitem{Muong-2:2023cdq}
{\scshape Muon g-2} collaboration, D.~P. Aguillard et~al., \emph{{Measurement
  of the Positive Muon Anomalous Magnetic Moment to 0.20~ppm}},
  \href{https://doi.org/10.1103/PhysRevLett.131.161802}{\emph{Phys. Rev. Lett.}
  {\bfseries 131} (2023) 161802}
  [\href{https://arxiv.org/abs/2308.06230}{{\ttfamily 2308.06230}}].

\bibitem{Aoyama:2012wk}
T.~Aoyama, M.~Hayakawa, T.~Kinoshita and M.~Nio, \emph{{Complete Tenth-Order
  QED Contribution to the Muon g-2}},
  \href{https://doi.org/10.1103/PhysRevLett.109.111808}{\emph{Phys. Rev. Lett.}
  {\bfseries 109} (2012) 111808}
  [\href{https://arxiv.org/abs/1205.5370}{{\ttfamily 1205.5370}}].

\bibitem{Aoyama:2019ryr}
T.~Aoyama, T.~Kinoshita and M.~Nio, \emph{{Theory of the Anomalous Magnetic
  Moment of the Electron}},
  \href{https://doi.org/10.3390/atoms7010028}{\emph{Atoms} {\bfseries 7} (2019)
  28}.

\bibitem{Czarnecki:2002nt}
A.~Czarnecki, W.~J. Marciano and A.~Vainshtein, \emph{{Refinements in
  electroweak contributions to the muon anomalous magnetic moment}},
  \href{https://doi.org/10.1103/PhysRevD.67.073006,
  10.1103/PhysRevD.73.119901}{\emph{Phys. Rev. D} {\bfseries 67} (2003) 073006}
  [\href{https://arxiv.org/abs/hep-ph/0212229}{{\ttfamily hep-ph/0212229}}].

\bibitem{Gnendiger:2013pva}
C.~Gnendiger, D.~St{\"o}ckinger and H.~St{\"o}ckinger-Kim, \emph{{The
  electroweak contributions to $(g-2)_\mu$ after the Higgs boson mass
  measurement}}, \href{https://doi.org/10.1103/PhysRevD.88.053005}{\emph{Phys.
  Rev. D} {\bfseries 88} (2013) 053005}
  [\href{https://arxiv.org/abs/1306.5546}{{\ttfamily 1306.5546}}].

\bibitem{Blum:2019ugy}
T.~Blum, N.~Christ, M.~Hayakawa, T.~Izubuchi, L.~Jin, C.~Jung et~al.,
  \emph{{Hadronic Light-by-Light Scattering Contribution to the Muon Anomalous
  Magnetic Moment from Lattice QCD}},
  \href{https://doi.org/10.1103/PhysRevLett.124.132002}{\emph{Phys. Rev. Lett.}
  {\bfseries 124} (2020) 132002}
  [\href{https://arxiv.org/abs/1911.08123}{{\ttfamily 1911.08123}}].

\bibitem{Chao:2021tvp}
E.-H. Chao, R.~J. Hudspith, A.~G\'erardin, J.~R. Green, H.~B. Meyer and
  K.~Ottnad, \emph{{Hadronic light-by-light contribution to $(g-2)_\mu $ from
  lattice QCD: a complete calculation}},
  \href{https://doi.org/10.1140/epjc/s10052-021-09455-4}{\emph{Eur. Phys. J. C}
  {\bfseries 81} (2021) 651}
  [\href{https://arxiv.org/abs/2104.02632}{{\ttfamily 2104.02632}}].

\bibitem{Chao:2022xzg}
E.-H. Chao, R.~J. Hudspith, A.~G\'erardin, J.~R. Green and H.~B. Meyer,
  \emph{{The charm-quark contribution to light-by-light scattering in the muon
  $(g-2)$ from lattice QCD}},
  \href{https://doi.org/10.1140/epjc/s10052-022-10589-2}{\emph{Eur. Phys. J. C}
  {\bfseries 82} (2022) 664}
  [\href{https://arxiv.org/abs/2204.08844}{{\ttfamily 2204.08844}}].

\bibitem{Blum:2023vlm}
T.~Blum, N.~Christ, M.~Hayakawa, T.~Izubuchi, L.~Jin, C.~Jung et~al.,
  \emph{{Hadronic light-by-light contribution to the muon anomaly from lattice
  QCD with infinite volume QED at physical pion mass}},
  \href{https://arxiv.org/abs/2304.04423}{{\ttfamily 2304.04423}}.

\bibitem{Brodsky:1967sr}
S.~J. Brodsky and E.~De~Rafael, \emph{{Suggested Boson-Lepton Pair Couplings
  and the Anomalous Magnetic Moment of the Muon}},
  \href{https://doi.org/10.1103/PhysRev.168.1620}{\emph{Phys. Rev.} {\bfseries
  168} (1968) 1620}.

\bibitem{Aoyama:2020ynm}
T.~Aoyama et~al., \emph{{The anomalous magnetic moment of the muon in the
  Standard Model}},
  \href{https://doi.org/10.1016/j.physrep.2020.07.006}{\emph{Phys. Rept.}
  {\bfseries 887} (2020) 1} [\href{https://arxiv.org/abs/2006.04822}{{\ttfamily
  2006.04822}}].

\bibitem{Davier:2017zfy}
M.~Davier, A.~Hoecker, B.~Malaescu and Z.~Zhang, \emph{{Reevaluation of the
  hadronic vacuum polarisation contributions to the Standard Model predictions
  of the muon $g-2$ and ${\alpha (m_Z^2)}$ using newest hadronic cross-section
  data}}, \href{https://doi.org/10.1140/epjc/s10052-017-5161-6}{\emph{Eur.
  Phys. J. C} {\bfseries 77} (2017) 827}
  [\href{https://arxiv.org/abs/1706.09436}{{\ttfamily 1706.09436}}].

\bibitem{Keshavarzi:2018mgv}
A.~Keshavarzi, D.~Nomura and T.~Teubner, \emph{{The muon $g-2$ and
  $\alpha(M_Z^2)$: a new data-based analysis}},
  \href{https://doi.org/10.1103/PhysRevD.97.114025}{\emph{Phys. Rev. D}
  {\bfseries 97} (2018) 114025}
  [\href{https://arxiv.org/abs/1802.02995}{{\ttfamily 1802.02995}}].

\bibitem{Colangelo:2018mtw}
G.~Colangelo, M.~Hoferichter and P.~Stoffer, \emph{{Two-pion contribution to
  hadronic vacuum polarization}},
  \href{https://doi.org/10.1007/JHEP02(2019)006}{\emph{JHEP} {\bfseries 02}
  (2019) 006} [\href{https://arxiv.org/abs/1810.00007}{{\ttfamily
  1810.00007}}].

\bibitem{Hoferichter:2019mqg}
M.~Hoferichter, B.-L. Hoid and B.~Kubis, \emph{{Three-pion contribution to
  hadronic vacuum polarization}},
  \href{https://doi.org/10.1007/JHEP08(2019)137}{\emph{JHEP} {\bfseries 08}
  (2019) 137} [\href{https://arxiv.org/abs/1907.01556}{{\ttfamily
  1907.01556}}].

\bibitem{Davier:2019can}
M.~Davier, A.~Hoecker, B.~Malaescu and Z.~Zhang, \emph{{A new evaluation of the
  hadronic vacuum polarisation contributions to the muon anomalous magnetic
  moment and to $\alpha(m_Z^2)$}},
  \href{https://doi.org/10.1140/epjc/s10052-020-7792-2}{\emph{Eur. Phys. J. C}
  {\bfseries 80} (2020) 241}
  [\href{https://arxiv.org/abs/1908.00921}{{\ttfamily 1908.00921}}].

\bibitem{Keshavarzi:2019abf}
A.~Keshavarzi, D.~Nomura and T.~Teubner, \emph{{The $g-2$ of charged leptons,
  $\alpha(M_Z^2)$ and the hyperfine splitting of muonium}},
  \href{https://doi.org/10.1103/PhysRevD.101.014029}{\emph{Phys. Rev. D}
  {\bfseries 101} (2020) 014029}
  [\href{https://arxiv.org/abs/1911.00367}{{\ttfamily 1911.00367}}].

\bibitem{Kurz:2014wya}
A.~Kurz, T.~Liu, P.~Marquard and M.~Steinhauser, \emph{{Hadronic contribution
  to the muon anomalous magnetic moment to next-to-next-to-leading order}},
  \href{https://doi.org/10.1016/j.physletb.2014.05.043}{\emph{Phys. Lett. B}
  {\bfseries 734} (2014) 144}
  [\href{https://arxiv.org/abs/1403.6400}{{\ttfamily 1403.6400}}].

\bibitem{CMD-3:2023alj}
{\scshape CMD-3} collaboration, F.~V. Ignatov et~al., \emph{{Measurement of the
  $e^+e^-\to\pi^+\pi^-$ cross section from threshold to 1.2 GeV with the CMD-3
  detector}},  \href{https://arxiv.org/abs/2302.08834}{{\ttfamily 2302.08834}}.

\bibitem{Abbiendi:2016xup}
G.~Abbiendi et~al., \emph{{Measuring the leading hadronic contribution to the
  muon g-2 via $\mu e$ scattering}},
  \href{https://doi.org/10.1140/epjc/s10052-017-4633-z}{\emph{Eur. Phys. J. C}
  {\bfseries 77} (2017) 139}
  [\href{https://arxiv.org/abs/1609.08987}{{\ttfamily 1609.08987}}].

\bibitem{Mohler:2017wnb}
D.~Mohler, S.~Schaefer and J.~Simeth, \emph{{CLS 2+1 flavor simulations at
  physical light- and strange-quark masses}},
  \href{https://doi.org/10.1051/epjconf/201817502010}{\emph{EPJ Web Conf.}
  {\bfseries 175} (2018) 02010}
  [\href{https://arxiv.org/abs/1712.04884}{{\ttfamily 1712.04884}}].

\bibitem{Blum:2018mom}
{\scshape RBC, UKQCD} collaboration, T.~Blum, P.~A. Boyle, V.~G{\"u}lpers,
  T.~Izubuchi, L.~Jin, C.~Jung et~al., \emph{{Calculation of the hadronic
  vacuum polarization contribution to the muon anomalous magnetic moment}},
  \href{https://doi.org/10.1103/PhysRevLett.121.022003}{\emph{Phys. Rev. Lett.}
  {\bfseries 121} (2018) 022003}
  [\href{https://arxiv.org/abs/1801.07224}{{\ttfamily 1801.07224}}].

\bibitem{Giusti:2019xct}
{\scshape European Twisted Mass} collaboration, D.~Giusti, V.~Lubicz,
  G.~Martinelli, F.~Sanfilippo and S.~Simula, \emph{{Electromagnetic and strong
  isospin-breaking corrections to the muon $g - 2$ from Lattice QCD+QED}},
  \href{https://doi.org/10.1103/PhysRevD.99.114502}{\emph{Phys. Rev. D}
  {\bfseries 99} (2019) 114502}
  [\href{https://arxiv.org/abs/1901.10462}{{\ttfamily 1901.10462}}].

\bibitem{Shintani:2019wai}
{\scshape PACS} collaboration, E.~Shintani and Y.~Kuramashi, \emph{{Hadronic
  vacuum polarization contribution to the muon $g-2$ with 2+1 flavor lattice
  QCD on a larger than (10 fm)$^4$ lattice at the physical point}},
  \href{https://doi.org/10.1103/PhysRevD.100.034517}{\emph{Phys. Rev. D}
  {\bfseries 100} (2019) 034517}
  [\href{https://arxiv.org/abs/1902.00885}{{\ttfamily 1902.00885}}].

\bibitem{Davies:2019efs}
{\scshape Fermilab Lattice, HPQCD, MILC} collaboration, C.~T.~H. Davies et~al.,
  \emph{{Hadronic-vacuum-polarization contribution to the
  muon\textquoteright{}s anomalous magnetic moment from four-flavor lattice
  QCD}}, \href{https://doi.org/10.1103/PhysRevD.101.034512}{\emph{Phys. Rev. D}
  {\bfseries 101} (2020) 034512}
  [\href{https://arxiv.org/abs/1902.04223}{{\ttfamily 1902.04223}}].

\bibitem{Gerardin:2019rua}
A.~G\'erardin, M.~C\`e, G.~von Hippel, B.~H\"orz, H.~B. Meyer, D.~Mohler
  et~al., \emph{{The leading hadronic contribution to $(g-2)_\mu$ from lattice
  QCD with $N_{\rm f}=2+1$ flavours of O($a$) improved Wilson quarks}},
  \href{https://doi.org/10.1103/PhysRevD.100.014510}{\emph{Phys. Rev. D}
  {\bfseries 100} (2019) 014510}
  [\href{https://arxiv.org/abs/1904.03120}{{\ttfamily 1904.03120}}].

\bibitem{Borsanyi:2020mff}
S.~Borsanyi et~al., \emph{{Leading hadronic contribution to the muon magnetic
  moment from lattice QCD}},
  \href{https://doi.org/10.1038/s41586-021-03418-1}{\emph{Nature} {\bfseries
  593} (2021) 51} [\href{https://arxiv.org/abs/2002.12347}{{\ttfamily
  2002.12347}}].

\bibitem{Lehner:2020crt}
C.~Lehner and A.~S. Meyer, \emph{{Consistency of hadronic vacuum polarization
  between lattice QCD and the R-ratio}},
  \href{https://doi.org/10.1103/PhysRevD.101.074515}{\emph{Phys. Rev. D}
  {\bfseries 101} (2020) 074515}
  [\href{https://arxiv.org/abs/2003.04177}{{\ttfamily 2003.04177}}].

\bibitem{Aubin:2022hgm}
C.~Aubin, T.~Blum, M.~Golterman and S.~Peris, \emph{{Muon anomalous magnetic
  moment with staggered fermions: Is the lattice spacing small enough?}},
  \href{https://doi.org/10.1103/PhysRevD.106.054503}{\emph{Phys. Rev. D}
  {\bfseries 106} (2022) 054503}
  [\href{https://arxiv.org/abs/2204.12256}{{\ttfamily 2204.12256}}].

\bibitem{Blum:2002ii}
T.~Blum, \emph{{Lattice calculation of the lowest order hadronic contribution
  to the muon anomalous magnetic moment}},
  \href{https://doi.org/10.1103/PhysRevLett.91.052001}{\emph{Phys. Rev. Lett.}
  {\bfseries 91} (2003) 052001}
  [\href{https://arxiv.org/abs/hep-lat/0212018}{{\ttfamily hep-lat/0212018}}].

\bibitem{Bernecker:2011gh}
D.~Bernecker and H.~B. Meyer, \emph{{Vector Correlators in Lattice QCD: Methods
  and applications}},
  \href{https://doi.org/10.1140/epja/i2011-11148-6}{\emph{Eur. Phys. J. A}
  {\bfseries 47} (2011) 148} [\href{https://arxiv.org/abs/1107.4388}{{\ttfamily
  1107.4388}}].

\bibitem{DellaMorte:2017dyu}
M.~Della~Morte, A.~Francis, V.~G{\"u}lpers, G.~Herdo{\'\i}za, G.~von Hippel,
  H.~Horch et~al., \emph{{The hadronic vacuum polarization contribution to the
  muon $g-2$ from lattice QCD}},
  \href{https://doi.org/10.1007/JHEP10(2017)020}{\emph{JHEP} {\bfseries 10}
  (2017) 020} [\href{https://arxiv.org/abs/1705.01775}{{\ttfamily
  1705.01775}}].

\bibitem{Colquhoun:2014ica}
{\scshape HPQCD} collaboration, B.~Colquhoun, R.~J. Dowdall, C.~T.~H. Davies,
  K.~Hornbostel and G.~P. Lepage, \emph{{$\Upsilon$ and $\Upsilon^{\prime}$
  Leptonic Widths, $a_{\mu}^b$ and $m_b$ from full lattice QCD}},
  \href{https://doi.org/10.1103/PhysRevD.91.074514}{\emph{Phys. Rev. D}
  {\bfseries 91} (2015) 074514}
  [\href{https://arxiv.org/abs/1408.5768}{{\ttfamily 1408.5768}}].

\bibitem{Bruno:2019nzm}
M.~Bruno, T.~Izubuchi, C.~Lehner and A.~S. Meyer, \emph{{Exclusive Channel
  Study of the Muon HVP}},
  \href{https://doi.org/10.22323/1.363.0239}{\emph{PoS} {\bfseries LATTICE2019}
  (2019) 239} [\href{https://arxiv.org/abs/1910.11745}{{\ttfamily
  1910.11745}}].

\bibitem{LehnerBounding2016}
C.~Lehner in \emph{{RBRC Workshop on Lattice Gauge Theories}}, 2016,
  \href{https://indico.bnl.gov/event/1628/contributions/2819/}{https://indico.bnl.gov/event/1628/contributions/2819/}.

\bibitem{Budapest-Marseille-Wuppertal:2017okr}
{\scshape Budapest-Marseille-Wuppertal} collaboration, S.~Borsanyi et~al.,
  \emph{{Hadronic vacuum polarization contribution to the anomalous magnetic
  moments of leptons from first principles}},
  \href{https://doi.org/10.1103/PhysRevLett.121.022002}{\emph{Phys. Rev. Lett.}
  {\bfseries 121} (2018) 022002}
  [\href{https://arxiv.org/abs/1711.04980}{{\ttfamily 1711.04980}}].

\bibitem{Paul:2023ksa}
S.~Paul, A.~D. Hanlon, B.~H\"orz, D.~Mohler, C.~Morningstar and H.~Wittig,
  \emph{{The long-distance behaviour of the vector-vector correlator from
  $\pi\pi$ scattering}}, \href{https://doi.org/10.22323/1.430.0073}{\emph{PoS}
  {\bfseries LATTICE2022} (2023) 073}.

\bibitem{Andersen:2018mau}
C.~Andersen, J.~Bulava, B.~H\"orz and C.~Morningstar, \emph{{The $I=1$
  pion-pion scattering amplitude and timelike pion form factor from $N_{\rm f}
  = 2+1$ lattice QCD}},
  \href{https://doi.org/10.1016/j.nuclphysb.2018.12.018}{\emph{Nucl. Phys. B}
  {\bfseries 939} (2019) 145}
  [\href{https://arxiv.org/abs/1808.05007}{{\ttfamily 1808.05007}}].

\bibitem{Neff:2001zr}
H.~Neff, N.~Eicker, T.~Lippert, J.~W. Negele and K.~Schilling, \emph{{On the
  low fermionic eigenmode dominance in QCD on the lattice}},
  \href{https://doi.org/10.1103/PhysRevD.64.114509}{\emph{Phys. Rev. D}
  {\bfseries 64} (2001) 114509}
  [\href{https://arxiv.org/abs/hep-lat/0106016}{{\ttfamily hep-lat/0106016}}].

\bibitem{Giusti:2004yp}
L.~Giusti, P.~Hernandez, M.~Laine, P.~Weisz and H.~Wittig, \emph{{Low-energy
  couplings of QCD from current correlators near the chiral limit}},
  \href{https://doi.org/10.1088/1126-6708/2004/04/013}{\emph{JHEP} {\bfseries
  04} (2004) 013} [\href{https://arxiv.org/abs/hep-lat/0402002}{{\ttfamily
  hep-lat/0402002}}].

\bibitem{DeGrand:2004qw}
T.~A. DeGrand and S.~Schaefer, \emph{{Improving meson two point functions in
  lattice QCD}}, \href{https://doi.org/10.1016/j.cpc.2004.02.006}{\emph{Comput.
  Phys. Commun.} {\bfseries 159} (2004) 185}
  [\href{https://arxiv.org/abs/hep-lat/0401011}{{\ttfamily hep-lat/0401011}}].

\bibitem{Bali:2009hu}
G.~S. Bali, S.~Collins and A.~Sch{\"a}fer, \emph{{Effective noise reduction
  techniques for disconnected loops in Lattice QCD}},
  \href{https://doi.org/10.1016/j.cpc.2010.05.008}{\emph{Comput. Phys. Commun.}
  {\bfseries 181} (2010) 1570}
  [\href{https://arxiv.org/abs/0910.3970}{{\ttfamily 0910.3970}}].

\bibitem{Blum:2012uh}
T.~Blum, T.~Izubuchi and E.~Shintani, \emph{{New class of variance-reduction
  techniques using lattice symmetries}},
  \href{https://doi.org/10.1103/PhysRevD.88.094503}{\emph{Phys. Rev. D}
  {\bfseries 88} (2013) 094503}
  [\href{https://arxiv.org/abs/1208.4349}{{\ttfamily 1208.4349}}].

\bibitem{Shintani:2014vja}
E.~Shintani, R.~Arthur, T.~Blum, T.~Izubuchi, C.~Jung and C.~Lehner,
  \emph{{Covariant approximation averaging}},
  \href{https://doi.org/10.1103/PhysRevD.91.114511}{\emph{Phys. Rev. D}
  {\bfseries 91} (2015) 114511}
  [\href{https://arxiv.org/abs/1402.0244}{{\ttfamily 1402.0244}}].

\bibitem{Ce:2016idq}
M.~C\`e, L.~Giusti and S.~Schaefer, \emph{{Domain decomposition, multi-level
  integration and exponential noise reduction in lattice QCD}},
  \href{https://doi.org/10.1103/PhysRevD.93.094507}{\emph{Phys. Rev. D}
  {\bfseries 93} (2016) 094507}
  [\href{https://arxiv.org/abs/1601.04587}{{\ttfamily 1601.04587}}].

\bibitem{Ce:2016ajy}
M.~C\`e, L.~Giusti and S.~Schaefer, \emph{{A local factorization of the fermion
  determinant in lattice QCD}},
  \href{https://doi.org/10.1103/PhysRevD.95.034503}{\emph{Phys. Rev. D}
  {\bfseries 95} (2017) 034503}
  [\href{https://arxiv.org/abs/1609.02419}{{\ttfamily 1609.02419}}].

\bibitem{DallaBrida:2020cik}
M.~Dalla~Brida, L.~Giusti, T.~Harris and M.~Pepe, \emph{{Multi-level Monte
  Carlo computation of the hadronic vacuum polarization contribution to
  $(g_\mu-2)$}},
  \href{https://doi.org/10.1016/j.physletb.2021.136191}{\emph{Phys. Lett. B}
  {\bfseries 816} (2021) 136191}
  [\href{https://arxiv.org/abs/2007.02973}{{\ttfamily 2007.02973}}].

\bibitem{Giusti:2018mdh}
D.~Giusti, F.~Sanfilippo and S.~Simula, \emph{{Light-quark contribution to the
  leading hadronic vacuum polarization term of the muon $g-2$ from twisted-mass
  fermions}}, \href{https://doi.org/10.1103/PhysRevD.98.114504}{\emph{Phys.
  Rev. D} {\bfseries 98} (2018) 114504}
  [\href{https://arxiv.org/abs/1808.00887}{{\ttfamily 1808.00887}}].

\bibitem{Aubin:2019usy}
C.~Aubin, T.~Blum, C.~Tu, M.~Golterman, C.~Jung and S.~Peris, \emph{{Light
  quark vacuum polarization at the physical point and contribution to the muon
  $g-2$}}, \href{https://doi.org/10.1103/PhysRevD.101.014503}{\emph{Phys. Rev.
  D} {\bfseries 101} (2020) 014503}
  [\href{https://arxiv.org/abs/1905.09307}{{\ttfamily 1905.09307}}].

\bibitem{Bijnens:2017esv}
J.~Bijnens and J.~Relefors, \emph{{Vector two-point functions in finite volume
  using partially quenched chiral perturbation theory at two loops}},
  \href{https://doi.org/10.1007/JHEP12(2017)114}{\emph{JHEP} {\bfseries 12}
  (2017) 114} [\href{https://arxiv.org/abs/1710.04479}{{\ttfamily
  1710.04479}}].

\bibitem{Aubin:2020scy}
C.~Aubin, T.~Blum, M.~Golterman and S.~Peris, \emph{{Application of effective
  field theory to finite-volume effects in $a_\mu^{\rm HVP}$}},
  \href{https://doi.org/10.1103/PhysRevD.102.094511}{\emph{Phys. Rev. D}
  {\bfseries 102} (2020) 094511}
  [\href{https://arxiv.org/abs/2008.03809}{{\ttfamily 2008.03809}}].

\bibitem{Sakurai:1960ju}
J.~J. Sakurai, \emph{{Theory of strong interactions}},
  \href{https://doi.org/10.1016/0003-4916(60)90126-3}{\emph{Annals Phys.}
  {\bfseries 11} (1960) 1}.

\bibitem{Jegerlehner:2011ti}
F.~Jegerlehner and R.~Szafron, \emph{{$\rho^0 - \gamma$ mixing in the neutral
  channel pion form factor $F_{\pi}^{(e)}(s)$ and its role in comparing $e^+
  e^-$ with $\tau$ spectral functions}},
  \href{https://doi.org/10.1140/epjc/s10052-011-1632-3}{\emph{Eur. Phys. J. C}
  {\bfseries 71} (2011) 1632}
  [\href{https://arxiv.org/abs/1101.2872}{{\ttfamily 1101.2872}}].

\bibitem{Chakraborty:2016mwy}
{\scshape HPQCD} collaboration, B.~Chakraborty, C.~T.~H. Davies, P.~G.
  de~Oliviera, J.~Koponen, G.~P. Lepage and R.~S. Van~de Water, \emph{{The
  hadronic vacuum polarization contribution to $a_{\mu}$ from full lattice
  QCD}}, \href{https://doi.org/10.1103/PhysRevD.96.034516}{\emph{Phys. Rev. D}
  {\bfseries 96} (2017) 034516}
  [\href{https://arxiv.org/abs/1601.03071}{{\ttfamily 1601.03071}}].

\bibitem{Luscher:1986pf}
M.~L{\"u}scher, \emph{{Volume Dependence of the Energy Spectrum in Massive
  Quantum Field Theories. 2. Scattering States}},
  \href{https://doi.org/10.1007/BF01211097}{\emph{Commun. Math. Phys.}
  {\bfseries 105} (1986) 153}.

\bibitem{Lellouch:2000pv}
L.~Lellouch and M.~L{\"u}scher, \emph{{Weak transition matrix elements from
  finite volume correlation functions}}, {\emph{Commun. Math. Phys.} {\bfseries
  219} (2001) 31} [\href{https://arxiv.org/abs/hep-lat/0003023}{{\ttfamily
  hep-lat/0003023}}].

\bibitem{Meyer:2011um}
H.~B. Meyer, \emph{{Lattice QCD and the Timelike Pion Form Factor}},
  \href{https://doi.org/10.1103/PhysRevLett.107.072002}{\emph{Phys. Rev. Lett.}
  {\bfseries 107} (2011) 072002}
  [\href{https://arxiv.org/abs/1105.1892}{{\ttfamily 1105.1892}}].

\bibitem{Gounaris:1968mw}
G.~J. Gounaris and J.~J. Sakurai, \emph{{Finite width corrections to the vector
  meson dominance prediction for $\rho \to e^+ e^-$}},
  \href{https://doi.org/10.1103/PhysRevLett.21.244}{\emph{Phys. Rev. Lett.}
  {\bfseries 21} (1968) 244}.

\bibitem{Francis:2013fzp}
A.~Francis, B.~J{\"a}ger, H.~B. Meyer and H.~Wittig, \emph{{A new
  representation of the Adler function for lattice QCD}},
  \href{https://doi.org/10.1103/PhysRevD.88.054502}{\emph{Phys. Rev. D}
  {\bfseries 88} (2013) 054502}
  [\href{https://arxiv.org/abs/1306.2532}{{\ttfamily 1306.2532}}].

\bibitem{Hansen:2019rbh}
M.~T. Hansen and A.~Patella, \emph{{Finite-volume effects in $(g-2)^{{\rm
  HVP,LO}}_\mu$}},
  \href{https://doi.org/10.1103/PhysRevLett.123.172001}{\emph{Phys. Rev. Lett.}
  {\bfseries 123} (2019) 172001}
  [\href{https://arxiv.org/abs/1904.10010}{{\ttfamily 1904.10010}}].

\bibitem{Hansen:2020whp}
M.~T. Hansen and A.~Patella, \emph{{Finite-volume and thermal effects in the
  leading-HVP contribution to muonic ($g-2$)}},
  \href{https://doi.org/10.1007/JHEP10(2020)029}{\emph{JHEP} {\bfseries 10}
  (2020) 029} [\href{https://arxiv.org/abs/2004.03935}{{\ttfamily
  2004.03935}}].

\bibitem{Ce:2021xgd}
M.~C\`e, T.~Harris, H.~B. Meyer, A.~Toniato and C.~T\"or\"ok, \emph{{Vacuum
  correlators at short distances from lattice QCD}},
  \href{https://doi.org/10.1007/JHEP12(2021)215}{\emph{JHEP} {\bfseries 12}
  (2021) 215} [\href{https://arxiv.org/abs/2106.15293}{{\ttfamily
  2106.15293}}].

\bibitem{Sommer:2022wac}
R.~Sommer, L.~Chimirri and N.~Husung, \emph{{Log-enhanced discretization errors
  in integrated correlation functions}},
  \href{https://doi.org/10.22323/1.430.0358}{\emph{PoS} {\bfseries LATTICE2022}
  (2023) 358} [\href{https://arxiv.org/abs/2211.15750}{{\ttfamily
  2211.15750}}].

\bibitem{Husung:2019ytz}
N.~Husung, P.~Marquard and R.~Sommer, \emph{{Asymptotic behavior of cutoff
  effects in Yang\textendash{}Mills theory and in Wilson\textquoteright{}s
  lattice QCD}},
  \href{https://doi.org/10.1140/epjc/s10052-020-7685-4}{\emph{Eur. Phys. J. C}
  {\bfseries 80} (2020) 200}
  [\href{https://arxiv.org/abs/1912.08498}{{\ttfamily 1912.08498}}].

\bibitem{Husung:2021mfl}
N.~Husung, P.~Marquard and R.~Sommer, \emph{{The asymptotic approach to the
  continuum of lattice QCD spectral observables}},
  \href{https://doi.org/10.1016/j.physletb.2022.137069}{\emph{Phys. Lett. B}
  {\bfseries 829} (2022) 137069}
  [\href{https://arxiv.org/abs/2111.02347}{{\ttfamily 2111.02347}}].

\bibitem{Husung:Lattice2023}
N.~Husung, \emph{{SymEFT predictions for local fermion bilinears}}, {\emph{PoS}
  {\bfseries LATTICE2023} (2024) 364}
  [\href{https://arxiv.org/abs/2401.04303}{{\ttfamily 2401.04303}}].

\bibitem{Ce:2022kxy}
M.~C\`e et~al., \emph{{Window observable for the hadronic vacuum polarization
  contribution to the muon g-2 from lattice QCD}},
  \href{https://doi.org/10.1103/PhysRevD.106.114502}{\emph{Phys. Rev. D}
  {\bfseries 106} (2022) 114502}
  [\href{https://arxiv.org/abs/2206.06582}{{\ttfamily 2206.06582}}].

\bibitem{RBC:2023pvn}
{\scshape RBC, UKQCD} collaboration, T.~Blum et~al., \emph{{Update of Euclidean
  windows of the hadronic vacuum polarization}},
  \href{https://doi.org/10.1103/PhysRevD.108.054507}{\emph{Phys. Rev. D}
  {\bfseries 108} (2023) 054507}
  [\href{https://arxiv.org/abs/2301.08696}{{\ttfamily 2301.08696}}].

\bibitem{ExtendedTwistedMass:2022jpw}
{\scshape Extended Twisted Mass} collaboration, C.~Alexandrou et~al.,
  \emph{{Lattice calculation of the short and intermediate time-distance
  hadronic vacuum polarization contributions to the muon magnetic moment using
  twisted-mass fermions}},
  \href{https://doi.org/10.1103/PhysRevD.107.074506}{\emph{Phys. Rev. D}
  {\bfseries 107} (2023) 074506}
  [\href{https://arxiv.org/abs/2206.15084}{{\ttfamily 2206.15084}}].

\bibitem{Aoki:2021kgd}
{\scshape Flavour Lattice Averaging Group (FLAG)} collaboration, Y.~Aoki
  et~al., \emph{{FLAG Review 2021}},
  \href{https://doi.org/10.1140/epjc/s10052-022-10536-1}{\emph{Eur. Phys. J. C}
  {\bfseries 82} (2022) 869}
  [\href{https://arxiv.org/abs/2111.09849}{{\ttfamily 2111.09849}}].

\bibitem{Colangelo:2022vok}
G.~Colangelo et~al., \emph{{Data-driven evaluations of Euclidean windows to
  scrutinize hadronic vacuum polarization}},
  \href{https://doi.org/10.1016/j.physletb.2022.137313}{\emph{Phys. Lett. B}
  {\bfseries 833} (2022) 137313}
  [\href{https://arxiv.org/abs/2205.12963}{{\ttfamily 2205.12963}}].

\bibitem{Boito:2022njs}
D.~Boito, M.~Golterman, K.~Maltman and S.~Peris, \emph{{Spectral-weight sum
  rules for the hadronic vacuum polarization}},
  \href{https://doi.org/10.1103/PhysRevD.107.034512}{\emph{Phys. Rev. D}
  {\bfseries 107} (2023) 034512}
  [\href{https://arxiv.org/abs/2210.13677}{{\ttfamily 2210.13677}}].

\bibitem{Davier:2023cyp}
M.~Davier, Z.~Fodor, A.~Gerardin, L.~Lellouch, B.~Malaescu, F.~M. Stokes
  et~al., \emph{{Hadronic vacuum polarization: comparing lattice QCD and
  data-driven results in systematically improvable ways}},
  \href{https://arxiv.org/abs/2308.04221}{{\ttfamily 2308.04221}}.

\bibitem{Colangelo:2022xfy}
G.~Colangelo, \emph{{Data-driven approaches to the evaluation of hadronic
  contributions to the $(g - 2)_\mu$}},
  \href{https://doi.org/10.1051/epjconf/202225801004}{\emph{EPJ Web Conf.}
  {\bfseries 258} (2022) 01004}.

\bibitem{ExtendedTwistedMassCollaborationETMC:2022sta}
{\scshape Extended Twisted Mass} collaboration, C.~Alexandrou et~al.,
  \emph{{Probing the Energy-Smeared $R$ Ratio Using Lattice QCD}},
  \href{https://doi.org/10.1103/PhysRevLett.130.241901}{\emph{Phys. Rev. Lett.}
  {\bfseries 130} (2023) 241901}
  [\href{https://arxiv.org/abs/2212.08467}{{\ttfamily 2212.08467}}].

\bibitem{Giusti:2021dvd}
D.~Giusti and S.~Simula, \emph{{Window contributions to the muon hadronic
  vacuum polarization with twisted-mass fermions}},
  \href{https://doi.org/10.22323/1.396.0189}{\emph{PoS} {\bfseries LATTICE2021}
  (2022) 189} [\href{https://arxiv.org/abs/2111.15329}{{\ttfamily
  2111.15329}}].

\bibitem{Wittig:2023pcl}
H.~Wittig, \emph{{Progress on $(g-2)_\mu$ from Lattice QCD}},  in \emph{{57th
  Rencontres de Moriond on Electroweak Interactions and Unified Theories}}, 6,
  2023, \href{https://arxiv.org/abs/2306.04165}{{\ttfamily 2306.04165}}.

\bibitem{Miranda:2020wdg}
J.~A. Miranda and P.~Roig, \emph{{New $\tau$-based evaluation of the hadronic
  contribution to the vacuum polarization piece of the muon anomalous magnetic
  moment}}, \href{https://doi.org/10.1103/PhysRevD.102.114017}{\emph{Phys. Rev.
  D} {\bfseries 102} (2020) 114017}
  [\href{https://arxiv.org/abs/2007.11019}{{\ttfamily 2007.11019}}].

\bibitem{Masjuan:2023qsp}
P.~Masjuan, A.~Miranda and P.~Roig, \emph{{$\tau$ data-driven evaluation of
  Euclidean windows for the hadronic vacuum polarization}},
  \href{https://arxiv.org/abs/2305.20005}{{\ttfamily 2305.20005}}.

\bibitem{Davier:2023fpl}
M.~Davier, A.~Hoecker, A.~M. Lutz, B.~Malaescu and Z.~Zhang, \emph{{Tensions in
  $e^+e^-\to\pi^+\pi^-(\gamma)$ measurements: the new landscape of data-driven
  hadronic vacuum polarization predictions for the muon $g-2$}},
  \href{https://arxiv.org/abs/2312.02053}{{\ttfamily 2312.02053}}.

\bibitem{Benton:2023dci}
G.~Benton, D.~Boito, M.~Golterman, A.~Keshavarzi, K.~Maltman and S.~Peris,
  \emph{{Data-Driven Determination of the Light-Quark Connected Component of
  the Intermediate-Window Contribution to the Muon g-2}},
  \href{https://doi.org/10.1103/PhysRevLett.131.251803}{\emph{Phys. Rev. Lett.}
  {\bfseries 131} (2023) 251803}
  [\href{https://arxiv.org/abs/2306.16808}{{\ttfamily 2306.16808}}].

\bibitem{Meyer:2017hjv}
H.~B. Meyer, \emph{{Lorentz-covariant coordinate-space representation of the
  leading hadronic contribution to the anomalous magnetic moment of the muon}},
  \href{https://doi.org/10.1140/epjc/s10052-017-5200-3}{\emph{Eur. Phys. J. C}
  {\bfseries 77} (2017) 616}
  [\href{https://arxiv.org/abs/1706.01139}{{\ttfamily 1706.01139}}].

\bibitem{Chao:2022ycy}
E.-H. Chao, H.~B. Meyer and J.~Parrino, \emph{{Coordinate-space calculation of
  the window observable for the hadronic vacuum polarization contribution to
  (g-2)$_{\mu}$}},
  \href{https://doi.org/10.1103/PhysRevD.107.054505}{\emph{Phys. Rev. D}
  {\bfseries 107} (2023) 054505}
  [\href{https://arxiv.org/abs/2211.15581}{{\ttfamily 2211.15581}}].

\bibitem{Francis:2019muy}
A.~Francis, P.~Fritzsch, M.~L\"uscher and A.~Rago, \emph{{Master-field
  simulations of O($a$)-improved lattice QCD: Algorithms, stability and
  exactness}}, \href{https://doi.org/10.1016/j.cpc.2020.107355}{\emph{Comput.
  Phys. Commun.} {\bfseries 255} (2020) 107355}
  [\href{https://arxiv.org/abs/1911.04533}{{\ttfamily 1911.04533}}].

\bibitem{Wang:2022lkq}
{\scshape chiQCD} collaboration, G.~Wang, T.~Draper, K.-F. Liu and Y.-B. Yang,
  \emph{{Muon g-2 with overlap valence fermions}},
  \href{https://doi.org/10.1103/PhysRevD.107.034513}{\emph{Phys. Rev. D}
  {\bfseries 107} (2023) 034513}
  [\href{https://arxiv.org/abs/2204.01280}{{\ttfamily 2204.01280}}].

\bibitem{FermilabLatticeHPQCD:2023jof}
{\scshape Fermilab Lattice, HPQCD, MILC} collaboration, A.~Bazavov et~al.,
  \emph{{Light-quark connected intermediate-window contributions to the muon
  g-2 hadronic vacuum polarization from lattice QCD}},
  \href{https://doi.org/10.1103/PhysRevD.107.114514}{\emph{Phys. Rev. D}
  {\bfseries 107} (2023) 114514}
  [\href{https://arxiv.org/abs/2301.08274}{{\ttfamily 2301.08274}}].

\bibitem{Borsanyi:2017zdw}
{\scshape Budapest-Marseille-Wuppertal} collaboration, S.~Borsanyi et~al.,
  \emph{{Hadronic vacuum polarization contribution to the anomalous magnetic
  moments of leptons from first principles}},
  \href{https://doi.org/10.1103/PhysRevLett.121.022002}{\emph{Phys. Rev. Lett.}
  {\bfseries 121} (2018) 022002}
  [\href{https://arxiv.org/abs/1711.04980}{{\ttfamily 1711.04980}}].

\bibitem{Gulpers:2014jaq}
V.~G\"ulpers, A.~Francis, B.~J\"ager, H.~Meyer, G.~von Hippel and H.~Wittig,
  \emph{{The leading disconnected contribution to the anomalous magnetic moment
  of the muon}}, \href{https://doi.org/10.22323/1.214.0128}{\emph{PoS}
  {\bfseries LATTICE2014} (2014) 128}
  [\href{https://arxiv.org/abs/1411.7592}{{\ttfamily 1411.7592}}].

\bibitem{DellaMorte:2010aq}
M.~Della~Morte and A.~J{\"u}ttner, \emph{{Quark disconnected diagrams in chiral
  perturbation theory}},
  \href{https://doi.org/10.1007/JHEP11(2010)154}{\emph{JHEP} {\bfseries 1011}
  (2010) 154} [\href{https://arxiv.org/abs/1009.3783}{{\ttfamily 1009.3783}}].

\bibitem{Stathopoulos:2013aci}
A.~Stathopoulos, J.~Laeuchli and K.~Orginos, \emph{{Hierarchical Probing for
  Estimating the Trace of the Matrix Inverse on Toroidal Lattices}},
  \href{https://doi.org/10.1137/120881452}{\emph{SIAM J. Sci. Comput.}
  {\bfseries 35} (2013) S299}
  [\href{https://arxiv.org/abs/1302.4018}{{\ttfamily 1302.4018}}].

\bibitem{Blum:2015you}
T.~Blum, P.~A. Boyle, T.~Izubuchi, L.~Jin, A.~J{\"u}ttner, C.~Lehner et~al.,
  \emph{{Calculation of the hadronic vacuum polarization disconnected
  contribution to the muon anomalous magnetic moment}},
  \href{https://doi.org/10.1103/PhysRevLett.116.232002}{\emph{Phys. Rev. Lett.}
  {\bfseries 116} (2016) 232002}
  [\href{https://arxiv.org/abs/1512.09054}{{\ttfamily 1512.09054}}].

\bibitem{Thron:1997iy}
C.~Thron, S.~J. Dong, K.~F. Liu and H.~P. Ying, \emph{{Pad{\'e} - Z(2)
  estimator of determinants}},
  \href{https://doi.org/10.1103/PhysRevD.57.1642}{\emph{Phys. Rev. D}
  {\bfseries 57} (1998) 1642}
  [\href{https://arxiv.org/abs/hep-lat/9707001}{{\ttfamily hep-lat/9707001}}].

\bibitem{McNeile:2006bz}
{\scshape UKQCD} collaboration, C.~McNeile and C.~Michael, \emph{{Decay width
  of light quark hybrid meson from the lattice}},
  \href{https://doi.org/10.1103/PhysRevD.73.074506}{\emph{Phys. Rev. D}
  {\bfseries 73} (2006) 074506}
  [\href{https://arxiv.org/abs/hep-lat/0603007}{{\ttfamily hep-lat/0603007}}].

\bibitem{Giusti:2019kff}
L.~Giusti, T.~Harris, A.~Nada and S.~Schaefer, \emph{{Frequency-splitting
  estimators of single-propagator traces}},
  \href{https://doi.org/10.1140/epjc/s10052-019-7049-0}{\emph{Eur. Phys. J. C}
  {\bfseries 79} (2019) 586}
  [\href{https://arxiv.org/abs/1903.10447}{{\ttfamily 1903.10447}}].

\bibitem{Ce:2022eix}
M.~C\`e, A.~G\'erardin, G.~von Hippel, H.~B. Meyer, K.~Miura, K.~Ottnad et~al.,
  \emph{{The hadronic running of the electromagnetic coupling and the
  electroweak mixing angle from lattice QCD}},
  \href{https://doi.org/10.1007/JHEP08(2022)220}{\emph{JHEP} {\bfseries 08}
  (2022) 220} [\href{https://arxiv.org/abs/2203.08676}{{\ttfamily
  2203.08676}}].

\bibitem{Patella:2017fgk}
A.~Patella, \emph{{QED Corrections to Hadronic Observables}},
  \href{https://doi.org/10.22323/1.256.0020}{\emph{PoS} {\bfseries LATTICE2016}
  (2017) 020} [\href{https://arxiv.org/abs/1702.03857}{{\ttfamily
  1702.03857}}].

\bibitem{DiCarlo:Lattice2023}
M.~Di~Carlo, \emph{{Isospin-breaking corrections to weak decays: the current
  status and a new infrared improvement}},
  \href{https://doi.org/10.22323/1.453.0120}{\emph{PoS} {\bfseries LATTICE2023}
  (2024) 120} [\href{https://arxiv.org/abs/2401.07666}{{\ttfamily
  2401.07666}}].

\bibitem{Westin:2020blm}
A.~Westin, W.~Kamleh, R.~Young, J.~Zanotti, R.~Horsley, Y.~Nakamura et~al.,
  \emph{{Anomalous magnetic moment of the muon with dynamical QCD+QED}},
  \href{https://doi.org/10.1051/epjconf/202024506035}{\emph{EPJ Web Conf.}
  {\bfseries 245} (2020) 06035}.

\bibitem{RCstar:2022yjz}
{\scshape RCstar} collaboration, L.~Bushnaq, I.~Campos, M.~Catillo,
  A.~Cotellucci, M.~Dale, P.~Fritzsch et~al., \emph{{First results on QCD+QED
  with C$^{*}$ boundary conditions}},
  \href{https://doi.org/10.1007/JHEP03(2023)012}{\emph{JHEP} {\bfseries 03}
  (2023) 012} [\href{https://arxiv.org/abs/2209.13183}{{\ttfamily
  2209.13183}}].

\bibitem{FermilabLattice:2017wgj}
{\scshape Fermilab Lattice, HPQCD, MILC} collaboration, B.~Chakraborty et~al.,
  \emph{{Strong-Isospin-Breaking Correction to the Muon Anomalous Magnetic
  Moment from Lattice QCD at the Physical Point}},
  \href{https://doi.org/10.1103/PhysRevLett.120.152001}{\emph{Phys. Rev. Lett.}
  {\bfseries 120} (2018) 152001}
  [\href{https://arxiv.org/abs/1710.11212}{{\ttfamily 1710.11212}}].

\bibitem{deDivitiis:2011eh}
G.~M. de~Divitiis et~al., \emph{{Isospin breaking effects due to the up-down
  mass difference in lattice QCD}},
  \href{https://doi.org/10.1007/JHEP04(2012)124}{\emph{JHEP} {\bfseries 04}
  (2012) 124} [\href{https://arxiv.org/abs/1110.6294}{{\ttfamily 1110.6294}}].

\bibitem{deDivitiis:2013xla}
{\scshape RM123} collaboration, G.~M. de~Divitiis, R.~Frezzotti, V.~Lubicz,
  G.~Martinelli, R.~Petronzio, G.~C. Rossi et~al., \emph{{Leading isospin
  breaking effects on the lattice}},
  \href{https://doi.org/10.1103/PhysRevD.87.114505}{\emph{Phys. Rev. D}
  {\bfseries 87} (2013) 114505}
  [\href{https://arxiv.org/abs/1303.4896}{{\ttfamily 1303.4896}}].

\bibitem{Biloshytskyi:2022ets}
V.~Biloshytskyi, E.-H. Chao, A.~G\'erardin, J.~R. Green, F.~Hagelstein, H.~B.
  Meyer et~al., \emph{{Forward light-by-light scattering and electromagnetic
  correction to hadronic vacuum polarization}},
  \href{https://doi.org/10.1007/JHEP03(2023)194}{\emph{JHEP} {\bfseries 03}
  (2023) 194} [\href{https://arxiv.org/abs/2209.02149}{{\ttfamily
  2209.02149}}].

\bibitem{Chao:2023lxw}
E.-H. Chao, H.~B. Meyer and J.~Parrino, \emph{{Coordinate-space calculation of
  QED corrections to the hadronic vacuum polarization contribution to
  $(g-2)_\mu$}}, \href{https://doi.org/10.22323/1.453.0256}{\emph{PoS}
  {\bfseries LATTICE2023} (2023) 256}
  [\href{https://arxiv.org/abs/2310.20556}{{\ttfamily 2310.20556}}].

\bibitem{Green:2015sra}
J.~Green, O.~Gryniuk, G.~von Hippel, H.~B. Meyer and V.~Pascalutsa,
  \emph{{Lattice QCD calculation of hadronic light-by-light scattering}},
  \href{https://doi.org/10.1103/PhysRevLett.115.222003}{\emph{Phys. Rev. Lett.}
  {\bfseries 115} (2015) 222003}
  [\href{https://arxiv.org/abs/1507.01577}{{\ttfamily 1507.01577}}].

\bibitem{Asmussen:2016lse}
N.~Asmussen, J.~Green, H.~B. Meyer and A.~Nyffeler, \emph{{Position-space
  approach to hadronic light-by-light scattering in the muon $g-2$ on the
  lattice}}, \href{https://doi.org/10.22323/1.256.0164}{\emph{PoS} {\bfseries
  LATTICE2016} (2016) 164} [\href{https://arxiv.org/abs/1609.08454}{{\ttfamily
  1609.08454}}].

\bibitem{Hayakawa:2008an}
M.~Hayakawa and S.~Uno, \emph{{QED in finite volume and finite size scaling
  effect on electromagnetic properties of hadrons}},
  \href{https://doi.org/10.1143/PTP.120.413}{\emph{Prog. Theor. Phys.}
  {\bfseries 120} (2008) 413}
  [\href{https://arxiv.org/abs/0804.2044}{{\ttfamily 0804.2044}}].

\bibitem{Bijnens:2019ejw}
J.~Bijnens, J.~Harrison, N.~Hermansson-Truedsson, T.~Janowski, A.~J\"uttner and
  A.~Portelli, \emph{{Electromagnetic finite-size effects to the hadronic
  vacuum polarization}},
  \href{https://doi.org/10.1103/PhysRevD.100.014508}{\emph{Phys. Rev. D}
  {\bfseries 100} (2019) 014508}
  [\href{https://arxiv.org/abs/1903.10591}{{\ttfamily 1903.10591}}].

\bibitem{Polley:1990tf}
L.~Polley and U.~J. Wiese, \emph{{Monopole condensate and monopole mass in U(1)
  lattice gauge theory}},
  \href{https://doi.org/10.1016/0550-3213(91)90380-G}{\emph{Nucl. Phys.}
  {\bfseries B356} (1991) 629}.

\bibitem{Lucini:2015hfa}
B.~Lucini, A.~Patella, A.~Ramos and N.~Tantalo, \emph{{Charged hadrons in local
  finite-volume QED+QCD with C$^{\ast}$ boundary conditions}},
  \href{https://doi.org/10.1007/JHEP02(2016)076}{\emph{JHEP} {\bfseries 02}
  (2016) 076} [\href{https://arxiv.org/abs/1509.01636}{{\ttfamily
  1509.01636}}].

\bibitem{Tantalo:2023onv}
N.~Tantalo, \emph{{Matching lattice QC+ED to Nature}},
  \href{https://doi.org/10.22323/1.430.0249}{\emph{PoS} {\bfseries LATTICE2022}
  (2023) 249} [\href{https://arxiv.org/abs/2301.02097}{{\ttfamily
  2301.02097}}].

\bibitem{Gerardin:2020gpp}
A.~G\'erardin, \emph{{The anomalous magnetic moment of the muon: status of
  lattice QCD calculations}},
  \href{https://doi.org/10.1140/epja/s10050-021-00426-7}{\emph{Eur. Phys. J. A}
  {\bfseries 57} (2021) 116}
  [\href{https://arxiv.org/abs/2012.03931}{{\ttfamily 2012.03931}}].

\bibitem{Ray:2022ycg}
G.~Ray et~al., \emph{{Calculating the QED correction to the hadronic vacuum
  polarisation on the lattice}},
  \href{https://doi.org/10.22323/1.430.0329}{\emph{PoS} {\bfseries LATTICE2022}
  (2023) 329} [\href{https://arxiv.org/abs/2212.12031}{{\ttfamily
  2212.12031}}].

\bibitem{Harris:2023zsl}
T.~Harris, V.~G\"ulpers, A.~Portelli and J.~Richings, \emph{{Efficiently
  unquenching QCD+QED at $\mathrm{O}(\alpha)$}},
  \href{https://doi.org/10.22323/1.430.0013}{\emph{PoS} {\bfseries LATTICE2022}
  (2023) 013} [\href{https://arxiv.org/abs/2301.03995}{{\ttfamily
  2301.03995}}].

\bibitem{Schaefer:2010hu}
{\scshape ALPHA} collaboration, S.~Schaefer, R.~Sommer and F.~Virotta,
  \emph{{Critical slowing down and error analysis in lattice QCD simulations}},
  \href{https://doi.org/10.1016/j.nuclphysb.2010.11.020}{\emph{Nucl. Phys. B}
  {\bfseries 845} (2011) 93} [\href{https://arxiv.org/abs/1009.5228}{{\ttfamily
  1009.5228}}].

\end{thebibliography}\endgroup
}

\end{document}